\documentclass[fleqn,usenatbib]{mnras}
\usepackage[T1]{fontenc}
\usepackage{ae,aecompl}
\usepackage{graphicx}
\usepackage{amsmath}
\usepackage{amssymb}
\usepackage{url}
\graphicspath{{figure/}}
\usepackage{soul} 
\usepackage{color}
\usepackage{xcolor}

\usepackage[normalem]{ulem}

\newcommand{\msun}{\mbox{${\rm M}_{\odot}$}}

\newcommand{\hkpc}{\ifmmode{h^{-1}{\rm kpc}}\;\else${h^{-1}}${\rm kpc}\fi}

\def\lesssim{\lower.5ex\hbox{$\; \buildrel < \over \sim \;$}}
\def\gtrsim{\lower.5ex\hbox{$\; \buildrel > \over \sim \;$}}

\begin{document}

\title[Mock Lightcones for CANDELS]{Mock Lightcones and Theory Friendly Catalogs for the CANDELS Survey}

\author[CANDELS] {
  Rachel S. Somerville$^{1,2}$\thanks{e-mail: rsomerville@flatironinstitute.org},
  Charlotte Olsen$^{2}$, L. Y. Aaron Yung$^{1,2}$, Camilla Pacifici$^3$, 
  \newauthor Henry C. Ferguson$^3$, Peter Behroozi$^4$, Shannon Osborne$^3$, Risa H. Wechsler$^5$,  \newauthor Viraj Pandya$^6$, 
  Sandra M. Faber$^6$, Joel R. Primack$^7$, Avishai Dekel$^8$ \\
$^1$Center for Computational Astrophysics, Flatiron Institute, 162 5th Avenue, New York, NY 10010\\
$^2$Department of Physics and Astronomy, Rutgers University, 136
Frelinghuysen Road, Piscataway, NJ 08854, USA\\
$^3$ Space Telescope Science Institute, 3700 San Martin Drive, Baltimore, MD 21218, USA\\
$^4$ Department of Astronomy and Steward Observatory, University of Arizona, Tucson, AZ 85721, USA\\
$^5$ Kavli Institute for Particle Astrophysics and Cosmology \& Physics Department, Stanford University, Stanford, CA 94305, USA; \\
SLAC National Accelerator Laboratory, Menlo Park, CA 94025, USA\\
$^6$ Department of Astronomy and Astrophysics, University of California, Santa Cruz, CA 95064, USA\\
$^7$ Physics Department, University of California, Santa Cruz, CA 95064, USA\\
$^8$ Racah Institute of Physics, The Hebrew University, Jerusalem 91904, Israel\\
}

\maketitle

\begin{abstract}
We present mock catalogs created to support the interpretation of the
CANDELS survey. We extract halos along past lightcones from the
Bolshoi Planck dissipationless N-body simulations and populate these
halos with galaxies using two different independently developed
semi-analytic models of galaxy formation and the empirical model
{\sc UniverseMachine}.  Our mock catalogs have geometries that encompass
the footprints of observations associated with the five CANDELS
fields. In order to allow field-to-field variance to be explored, we
have created eight realizations of each field. In this paper, we
present comparisons with observable global galaxy properties, including counts
in observed frame bands, luminosity functions, color-magnitude
distributions and color-color distributions. We additionally present
comparisons with physical galaxy parameters derived from SED fitting
for the CANDELS observations, such as stellar masses and star
formation rates. We find relatively good agreement between the model predictions and CANDELS observations for luminosity and stellar mass functions. We find poorer agreement for colors and star formation rate distributions. All of the mock lightcones as well as
curated ``theory friendly'' versions of the observational CANDELS
catalogs are made available through a web-based data hub.
\end{abstract}

\begin{keywords}
galaxies: formation, evolution, stellar content, high-redshift -- astronomical data base: surveys  
\end{keywords}

\section{Introduction}
\label{sec:intro}

The Cosmic Assembly Near-IR Deep Extragalactic Legacy Survey (CANDELS)
is a multi-cycle treasury program on the Hubble Space Telescope
\citep[HST; ][]{grogin:2011,koekemoer:2011}. The CANDELS
project\footnote{candels.ucolick.org} surveyed five widely separated
fields, each $\sim0.25$ square degrees, building on the legacy of
previous surveys such as the Great Observatories Origins Deep Survey
\citep[GOODS; ][]{giavalisco:2004}, the Hubble Ultra Deep Field
\citep[HUDF; ][]{beckwith:2006}, COSMOS \citep[AEGIS;
][]{scoville:2007}, and the UKIDSS Ultra Deep Survey \citep[UDS;
][]{cirasuolo:2007}. The major new contribution from CANDELS is
Near-IR imaging with the Wide Field Camera 3 (WFC3) in a ``wedding
cake'' configuration, with deeper imaging over two smaller areas and
shallower imaging over five wider areas. CANDELS-deep is sensitive
enough to reveal galaxy candidates viewed during ``cosmic dawn'' at
redshifts of $z\sim 6$--9. CANDELS-wide allows structural and
morphological properties of galaxies to be measured in the rest-frame
optical back to ``cosmic high noon'', $z\sim 2$--3, at the peak of
cosmic star formation activity. The HST component of CANDELS is
supplemented by a rich set of ancillary data from the UV through the
radio, including observations from the Chandra Space Telescope, the
Spitzer Space Telescope, the Herschel Space telescope, and numerous
ground-based facilities. 

An important goal of the CANDELS project is to provide to the
community a legacy database of high level data products such as object
catalogs, photometric redshifts, rest-frame photometry, and estimates
of physical parameters such as stellar masses and star formation
rates.  Because each CANDELS field has a different set of ancillary
data, the catalogs are documented in separate papers
\citep{guo:2013,galametz:2013,stefanon:2017,nayyeri:2017,barro:2019}. Photometric
redshift and stellar mass estimates have been presented in
\citet{dahlen:2013}, \citet{mobasher:2015}, and \citet{santini:2015},
and star formation rate estimates are presented in \citet{barro:2019}.

One goal of this paper is to document a set of ``theory friendly''
CANDELS high-level science products, which we have curated in order to
make it easier to compare the CANDELS results with the predictions of
theoretical models (or with other surveys). The ``theory friendly''
catalogs (hereafter TF-CANDELS catalogs) have a standard format, with
the same set of observational and derived quantities included, and have had a fairly generic set of data quality cuts pre-applied. The quantities included in the TF-CANDELS catalogs have been selected to comprise those that we expect to be of the most interest for comparison with theoretical models.

Another important component of the CANDELS project has been the
development of custom theoretical models and simulations to aid in the
interpretation of CANDELS results. One major part of the theory effort
has been the development of detailed ``mock catalogs'' tailored to the
characteristics of the CANDELS survey. To build these mock catalogs,
we have extracted ``lightcones'' from a large dissipationless N-body
simulation, with geometries matched to the five CANDELS fields. The
lightcones are lists of the masses, redshifts, and positions on the
sky (right ascension and declination) of halos extracted along a past
lightcone. We can then construct ``merger trees'' which describe the
build-up of these halos over time via merging of smaller halos. The
observable properties of the galaxies that form in these halos can
then be computed using an approach known as semi-analytic modeling.

Semi-analytic models (SAMs) of galaxy formation are a widely used tool
for studying the formation and evolution of galaxies in a cosmological
context. In this approach, one tracks bulk quantities such as diffuse
hot gas, cold star forming gas, stars, heavy elements, etc, using
approximations and phenomenological recipes. These models are set
within the backbone of the dark matter halo merger trees mentioned
above, which track the build-up of gravitationally collapsed
structures. They typically include modeling the shock heating and
radiative cooling of gas, star formation and stellar feedback,
chemical evolution, and morphological transformation via mergers. Some
recent models also include the formation and growth of supermassive
black holes and feedback from Active Galactic Nuclei (AGN). The
resulting star formation and chemical enrichment histories can then be
combined with stellar population models \citep[e.g.][]{Bruzual:2003}
and a treatment of attenuation by dust in order to obtain estimates of
luminosities at UV-NIR wavelengths.

Semi-analytic models adopt many simplifications and approximations,
and do not provide information that is as detailed as the output from
a numerical hydrodynamic simulation. But SAMs have the advantage of
much greater computational efficiency, as well as
flexibility. Moreover, numerical simulations must still adopt
phenomenological treatments of ``sub-grid physics'' to describe
physical processes that occur at scales smaller than the resolution of
the simulation \citep{SD15,naab:2017}. In many cases,
these recipes are similar to those utilized in SAMs. Modern SAMs and
cosmological numerical hydrodynamic simulations apparently yield very
consistent results, at least for many key global quantities \citep{SD15}.

An alternative method of linking dark matter halo properties with
observables is to use empirical models such as sub-halo abundance
matching models (SHAMs) or their variants \citep[see][for a recent
  review]{wechsler:2018}. Rather than attempting to implement {\it a
  priori} all of the detailed physical processes associated with
galaxy formation, these models derive mappings between dark matter
halo properties and observationally derived quantities, such that
observational constraints are satisfied. The {\sc UniverseMachine} developed
by \citet{Behroozi:2019} is an example of such an approach.

We have created a set of mock catalogs based on the CANDELS lightcones
using three different approaches: the Santa Cruz SAM developed by
R. Somerville and collaborators \citep[][and references
  therein]{somerville:2015}, the SAM code of Y. Lu and collaborators
\citep{Lu:2011}, and the {\sc UniverseMachine} \citep{Behroozi:2019}. In
\citet{Lu:2014}, we conducted an extensive comparison of the
predictions of the SC and Lu SAMs, as well as a third SAM by
\citet{Croton:2006}, for ``intrinsic'' galaxy properties over the
redshift range $z\sim 0$--6, such as stellar mass functions, the
stellar mass versus star formation rate and the fraction of quiescent
galaxies, cold gas fraction versus stellar mass, the mass-metallicity
relation, and the outflow rates of gas expelled by stellar
feedback. The three models were run in the same merger trees and were
all calibrated to reproduce the $z=0$ stellar mass function. Overall,
we found that the models produced fairly similar results, although
with some significant differences particularly at the highest
redshifts investigated. However, we did not compare the model predictions with actual CANDELS data in that work, as the high level data products were not yet available. 

The goals of this paper are three-fold: first, we document the details
of the construction and contents of the mock catalogs, which have
already been used in a number of CANDELS papers, and which we now
release to the community. Second, we present the predictions of the SC and Lu
SAMs for standard quantities such as observed counts,
rest-frame luminosity functions, color-magnitude relations, and
color-color diagrams.  We focus here on the redshift range $0.5
\lesssim z \lesssim 3$, which encompasses CANDELS ``cosmic noon''
science results. We compare the predictions of the SC SAM with the higher redshift
Universe $z\gtrsim 4$ in other work \citep{Yung:2019a,Yung:2019b}. Third, we document and release the new ``theory friendly" versions of the CANDELS observational catalogs. 

The structure of this paper is as follows. In \S\ref{sec:mockbuilding} we
describe how we created the mock catalogs, including providing
background on the underlying N-body simulations, the method used to
extract the lightcones, and some brief background about the models
that are used to predict galaxy properties.  In \S\ref{sec:tfcandels}
we briefly describe the CANDELS observations and the new ``theory friendly''
catalogs. In \S\ref{sec:results}, we present a comparison of the
predictions of the models with observed and derived quantities from CANDELS. We discuss our results, including a comparison with previous
work, in \S\ref{sec:discussion} and summarize and conclude in
\S\ref{sec:conclusions}. Throughout, we adopt the cosmological
parameters consistent with the recent analysis of the Planck survey
(as given in \S\ref{sec:sims}) and a Chabrier stellar initial mass
function \citep{Chabrier:2003}. All magnitudes are quoted in the AB
system.

\section{Building the Mock Catalogs}
\label{sec:mockbuilding}

\begin{figure} 
\begin{center}
\includegraphics[width=0.45\textwidth]{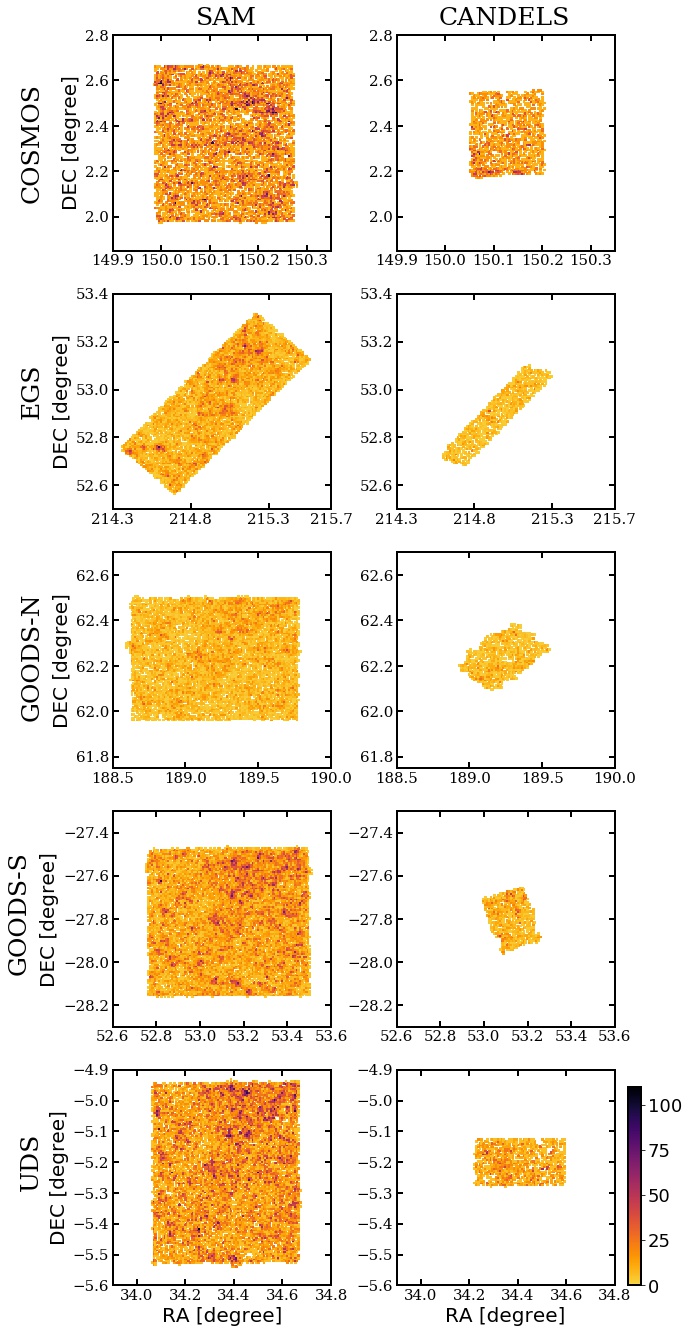}
\end{center}
\caption{Approximate footprints of the five mock lightcones (left), and the observed CANDELS fields, for a slice $0.9 < z < 1.1$ with F160W$<26$. The color scale shows the density of galaxies on the sky in arcmin$^{-2}$. The mock lightcones subtend a much larger area than the CANDELS HST footprint, by design. 
\label{fig:fields}}
\end{figure}

\begin{table}
\begin{center}
\centering
\begin{tabular}{lcc}
  \hline\hline
  field & dimensions (arcmin) & area (arcmin$^2$)\\
  \hline
  COSMOS & $17 \times 41$ & 697 \\
  EGS & $17 \times 46$ & 782 \\
  GOODS-N & $32 \times 32$ & 1024 \\
  GOODS-S & $39 \times 41$ & 1599\\
  UDS & $36 \times 35$ & 1260\\
  \hline
\end{tabular}
\caption{Dimensions and areas of the mock lightcones for the five CANDELS fields. Note that these dimensions are typically different from the HST footprint of the observed CANDELS fields.}
\label{tab:mocks}
\end{center}
\end{table}

\subsection{Simulations and Lightcones}
\label{sec:sims}
The lightcones used to construct our mock catalogs are extracted from
the Bolshoi Planck (hereafter BolshoiP) 
N-body simulations \citep{klypin:2014,rodriguez-puebla:2016}. The cosmological parameters are: matter density
$\Omega_m=0.307$, baryon density $\Omega_b=0.048$, Hubble parameter
$H_0=67.8$ km s$^{-1}$ Mpc$^{-1}$, tilt $n_s=0.96$, power spectrum normalization
$\sigma_8=0.823$. The BolshoiP simulation box is 369 (250 $h^{-1}$)
comoving Mpc on a side, with particle mass $2.2 \times 10^8$ \msun\
 ($1.5 \times 10^8$ $h^{-1}$ \msun), and a force resolution of 1.5 kpc
(1 $\hkpc$) in physical units.

Dark matter halos and subhalos were identified using the {\sc
  ROCKSTAR} code \citep{rockstar}. The halo catalogs are complete
above a mass of $\simeq 2.2 \times 10^{10} \msun$ (50 km s$^{-1}$). Merger trees have been constructed from these halo catalogs using the
{\sc Consistent Trees} code \citep{behroozi:ct}. All results presented here make use of
the halo virial mass definition of \citet{bryan:1998}, given in Eqn.~1 of \citet{rodriguez-puebla:2016}.

Lightcones are extracted from 164 snapshots between redshifts
$0-10$.  Lightcone origins and orientations are chosen randomly within the simulation volume.  The simulation has been constructed with periodic boundary conditions, and the simulated volume is replicated and tiled in all directions. Halos are collected along each lightcone from the snapshot closest to their cosmological redshift.  As CANDELS is comprised of pencil-beam surveys, no restrictions on sampling overlapping regions of the simulation volume are applied, as overlaps typically occur at redshift spacings $\Delta z > 1$.  The source code for
creating lightcones is available online\footnote{\url{https://bitbucket.org/pbehroozi/universemachine/src/master/}} in the \texttt{lightcone} package, and the full description of the algorithm is in \citet{behroozi:2020}. For more information on how to use the \texttt{lightcone} package, please see the online documention at \url{https://bitbucket.org/pbehroozi/universemachine/src/master/README.md\#markdown-header-making-new-lightcones}.

The geometry of each field is chosen to encompass the largest
footprint of any dataset used by the CANDELS project --- these are in
general significantly larger than the area of the HST mosaic.
Figure~\ref{fig:fields} shows the footprint of the mock catalogs in RA
and DEC on the sky compared with the F160W footprint of the actual
CANDELS fields.  The dimensions for each field are given in
Table~\ref{tab:mocks}. In order to investigate field-to-field
variance, we have created eight realizations of each field. Note that the lightcones used for the SC SAM and the Lu-SAM are for the same lines of sight, and contain identical sets of halos. The lightcones used for the {\sc UniverseMachine} are different lines of sight and contain different halos, but comprise a statistical representation of the same halo population. 

\subsection{Semi-Analytic Models}
\label{sec:sams}

The two semi-analytic models used in this work contain a similar suite
of physical processes, but these processes are parameterized and
implemented in different ways. Both models are based on merger trees,
which describe how dark matter halos collapse and merge to form larger
structures over time. The models contain prescriptions describing the
cosmological accretion of gas into halos, cooling of hot halo gas into
the interstellar medium (ISM) of galaxies, and the formation of stars
from cold ISM gas. In addition, the models track the return of mass
and metals to the ISM from massive stars and supernovae, and contain a
schematic treatment of ``stellar feedback'', the ejection of mass and
metals by stellar and supernova driven winds. The Santa Cruz model
contains a prescription for the formation and growth of supermassive
black holes, and associated ``black hole feedback'', while the Lu
models include a phenomenological halo-based quenching model. Both
models track the stellar mass in a ``disk'' and ``spheroid'' component
of each galaxy separately, allowing for simplified estimates of galaxy
morphology to be made. Both models additionally contain estimates of
the radial size of the disk component of each galaxy. The Santa Cruz
model also includes estimates of the size of the spheroid component,
based on the models developed by \citet{Porter:2014}. Note that the
version of the Santa Cruz models used here contains tracking of
multiphase gas and a molecular hydrogen based star formation recipe,
as described in \citet{somerville:2015}. The model parameters have
also been updated relative to those presented in
\citet{somerville:2015} to account for the BolshoiP cosmology
\citep[see][for details]{Yung:2019a}. In addition, as in \citet{Yung:2019a}, the filtering mass for photoionization squelching has been updated to the results from \citet{Okamoto:2008}. 

An important difference between the models is that the Lu models utilize the merger trees extracted directly from the BolshoiP simulations, and therefore the mass resolution is limited to $\sim 10^{11} \msun$ for root halos. The Santa Cruz SAMs use the ``root halos'' along the lightcones from BolshoiP, but construct the halo merger histories using the Extended Press-Schechter formalism as presented in \citet{somerville:2008}. Therefore, the halo merger histories depend only on halo mass and redshift, and do not carry a second-order dependence on the large scale environment. However, this means that the Santa Cruz mocks extend an order of magnitude further down in mass resolution, to root halos of $\sim 10^{10} \msun$.

Both SAMs carry out stellar population synthesis by combining the
predicted star formation and chemical enrichment histories with simple
stellar population models and analytic estimates of the effects of
dust attenuation, to predict galaxy spectral energy distributions. The
Santa Cruz models additionally utilize dust emission templates to
extend the SED predictions to longer wavelengths, where the light is
dominated by dust emission rather than starlight.  More details are
given in \S\ref{sec:stellpop} below.

For a detailed description of the semi-analytic models used in this
work, please refer to \citet[][hereafter L14]{Lu:2014} and references
therein, especially \citet{Lu:2011}, \citet[][S08]{somerville:2008},
\citet{somerville:2012}, \citet{Porter:2014}, and \citet{somerville:2015}.

\subsubsection{Substructure and Orphans}

Sub-halos are halos that have become subsumed within another
virialized halo. In the typical terminology of SAMs, sub-halos are
said to host ``satellite'' galaxies. Sub-halos are tidally stripped as
they orbit within their host halo. They may be tidally destroyed
before they merge, or they may merge with the central galaxy or with
another satellite. The SAMs used in this study treat sub-halos (which
host satellite galaxies) in different ways. The {\sc ROCKSTAR} catalogs
provide merger trees for sub-halos as well as distinct halos. However,
as with any simulation, the ability to explicitly track the evolution
of sub-halos is limited by the mass and force resolution of the
simulation \citep{vandenbosch:2018a,vandenbosch:2018b}. Moreover, the presence of baryons can affect the timescale for tidal stripping and destruction of satellites \citep[e.g.][]{garrison-kimmel:2017}, yet these effects are not accounted for self-consistently as our merger trees
are based on dark-matter only simulations.  Sub-halos that can no
longer be identified in the N-body outputs, but which may still have
surviving satellite galaxies associated with them, are commonly
referred to as ``orphans''. Many SAMs utilize semi-analytic recipes to
continue to track the evolution of orphans until they merge or are
tidally destroyed. For a detailed discussion of these issues, and a state of the art semi-analytic treatment of sub-halo evolution, see \citet{jiang:2020}.

The Santa Cruz SAM treats all satellite galaxies as ``orphans'' from
the time that they enter the host halo. A modified version of the
Chandrasekhar equation, which tracks the loss of orbital angular
momentum due to dynamical friction against the dark matter halo, is
used to estimate the radial distance of the satellite from the center
of the host halo as a function of time \citep{Boylan-Kolchin:2008}. As
the satellite orbits, a fixed amount of its mass is stripped off in
each orbit, following \citet{taylor:2001}. If the sub-halo's mass
drops below $M(<f_{\rm strip} r_s)$, where $f_{\rm strip}$ is an
adjustable parameter and $r_s$ is the Navarro-Frenk-White \citep{NFW}
scale radius, then the sub-halo is considered tidally destroyed. Its
stars are added to the ``diffuse stellar halo'' and its cold gas is
added to the hot gas reservoir. If the satellite survives until it
reaches the center of the halo, then the satellite is merged with the
central galaxy (satellites are not allowed to merge with other
satellites). The details of the treatment of mergers are described in
S08 and L14.

The Lu SAM uses the sub-halo information from the N-body catalogs to
follow the satellite population for as long as the sub-halo can be
resolved. When the sub-halo disappears from the N-body merger tree
catalog, its properties when it was last identified are used in a
formula that computes the dynamical friction time using the Chandrasekhar
formula as given in \citet{Binney-Tremaine}. The orphan satellite
is assumed to merge with the central galaxy after this time has
ellapsed. Tidal stripping and destruction of orphan satellites is not
accounted for. See L14 section A.7.3 for details.

\subsubsection{Stellar Populations and Dust}
\label{sec:stellpop}
Each semi-analytic model produces a prediction for the joint
distribution of ages and metallicities in each galaxy along the
lightcone at its observation time. These are obtained from the star
formation and chemical enrichment histories of all progenitors that
have merged into that galaxy by the output time. These age-metallicity
distributions are then convolved with stellar population synthesis
models to obtain intrinsic (non-dust-attenuated) spectral energy
distributions (SED) which may be convolved with any desired filter
response functions. Both SAMs use the stellar population
synthesis models of \citet[][BC03]{Bruzual:2003} with the
Padova 1994 isochrones and a Chabrier IMF. Note that the synthetic
SEDs currently do not currently include nebular emission.

If we write the mass of stars formed in all progenitors of a given
galaxy with ages between $t$, $t+dt$ and metallicities between $Z$ and
$Z+dZ$ as $\Psi(t, Z)\, dt\, dZ$, then the SED of the galaxy is
obtained by summing the ``simple stellar population'' components
provided by BC03 over all ages and metallicities:

\[ F_\lambda(t_{\rm obs}) = \int^{t_{\rm obs}}_{t_0} \int^{Z_{\rm  max}}_{Z_{\rm min}} T_{\rm dust}(\lambda) \Psi(t, Z) S_\lambda(t, Z) dt\, dZ \]

where in practice, the SSPs are provided at a set of discrete ages and
metallicities (196 ages and 6 metallicities, in the case of the BC03
models) so the integral is actually a sum. The timestep in the SAM is
chosen such that the time binning is at least as fine as that in
$S_\lambda(t, Z)$ at any point.

Dust attenuation is included through the term $T_{\rm dust}(\lambda)$,
which is given by
$T_{\rm dust}(\lambda) = 10.0^{-0.4 A_V k_{\lambda}}$
where $A_V$ is the attenuation in the rest-V band and $k_{\lambda}$ is
the attenuation as a function of wavelength relative to the $V$-band.

We model the rest V-band optical depth using the expressions:
\begin{gather*}
N_{\rm H} = m_{\rm{cold}}/(r_{\rm{gas}})^2 \\
\tau_{V,0} = f_{\rm dust}(z)\, \tau_{\rm{dust,0}}\, (Z_{\rm{cold}})^{\alpha_{\rm dust}} \, (N_{\rm H})^{\beta_{\rm dust}}
\end{gather*}
where $\tau_{\rm{dust,0}}$, $\alpha_{\rm dust}$, and $\beta_{\rm dust}$ are free parameters,
$Z_{\rm{cold}}$ is the metallicity of the cold gas,
$m_{\rm{cold}}$ is the mass of the cold gas in the disc, and
$r_{\rm{gas}}$ is the radius of the cold gas disc, which is
assumed to be a fixed multiple of the stellar scale length (see
S08). We adopt $\tau_{\rm{dust,0}}=0.2$, $\alpha_{\rm dust}=0.4$, and $\beta_{\rm dust}=1.0$.

Several works have found \citep[see e.g.][and references therein]{somerville:2012} that adopting this simple prescription with a fixed value of $\tau_{V,0}$ results in attenuation that is too strong at high redshift. As a result, we adopt an empirical redshift dependent functional form for $\tau_{V,0}$. For $z<3.5$, we adopt the redshift dependent correction factor
\begin{equation}
f_{\rm dust}(z) = (1+z)^{\gamma_{\rm dust}} \,
\end{equation}
and for $z>3.5$, we adopt the expression given in Section 2.4 of \citet{Yung:2019a}. This empirical relation was adjusted by hand to achieve a reasonable ``by-eye'' match to the observed rest-frame UV, B and V-band luminosity function from $z\sim 0$--4, and the observed rest-frame UV luminosity function at $z\gtrsim 4$. 

To compute the attenuation we assign a random inclination
to each disc galaxy and use a standard `slab' model; i.e. the extinction in
the $V$-band for a galaxy with inclination $i$ is given by:
\begin{equation}
A_V = -2.5 \log_{10}\left[\frac{1-\exp[-\tau_{V,0}/\cos(i)]}{\tau_{V,0}/\cos(i)}\right].
\end{equation}

For $k_{\lambda}$, we adopt the starburst attenuation curve of
\citet{Calzetti:2000}. We have also experimented with a two-component
(cirrus plus birthcloud) model for the attenuation, as presented in
S12. However, we found that the simpler Calzetti attenuation curve
does a better job of reproducing the colors of observed CANDELS
galaxies over the whole redshift range that we study here.

Dust emission modeling is included in the SC SAMs using the same approach described in S12, but adopting the \citet{chary:2001} emission templates. 

\begin{table}
	\centering
	\caption{Summary of recalibrated SC SAM parameters.}
	\label{tab:param}
	\begin{tabular}{ l  l  c } 
		\hline
		Parameter       & Description              & Value \\ [0.5ex] 
		\hline
		$\epsilon_\text{SN}$ &  SN feedback efficiency  & 1.7   \\ 
		$\alpha_\text{rh}$ & SN feedback slope & 3.0 \\
                $V_{\rm eject}$ & halo gas ejection scale & 130 km/s \\
		$\tau_{*,0}$ &  SF timescale normalization & 1.0   \\ 
		$y$ & Chemical yield (in solar units) & 2.0\\
		$\kappa_\text{AGN}$ & Radio mode AGN feedback & $3.0 \times 10^{-3}$ \\
		\hline
	\end{tabular}
\end{table}

\subsubsection{Calibration}
All cosmological models of galaxy formation contain parameterized
recipes, which are typically adjusted by tuning them to match a
selected subset of observations. For a detailed summary of the tunable
parameters in the three SAMs presented here, and the approach used to
tune them, please see L14. 
Some parameters were re-tuned relative to
the values used in L14, due to the change in cosmological parameters
from the original Bolshoi simulations (used in L14) to
BolshoiP. Table~\ref{tab:param} provides a summary of
parameters for the Santa Cruz SAM that have different values from those specified in \citet{somerville:2015}. Please see \citet{somerville:2015} for a full description of the parameters, and Table~1 in that work for a complete table of parameter values. The observations used for the calibration and the results of the calibration comparison are shown in \citet{Yung:2019a} Appendix~B. The calibration quantities include the stellar mass function, the stellar mass vs. cold gas fraction, stellar mass vs. metallicity relation, and the bulge mass vs. black hole mass relation.

\begin{table*}
\begin{center}
\centering
\begin{tabular}{lccccccc}
  \hline\hline
  field & reference & $\sigma$ & aperture & depth (W/D/UD) & RA & DEC & effective area \\
   & & & [arcsec] & [AB mag]& [degree]& [degree] & [arcmin$^2$]\\
  \hline
  COSMOS & \citet{nayyeri:2017} & 5 & 0.17 & 27.56& 150.116321 & +62.238572 & 216 \\
  EGS & \citet{stefanon:2017} & 5 & 0.20 & 27.6 & 214.825000 & +52.825000 & 198.6\\
  GOODS-N & \citet{barro:2019} & 5 & 0.17 & 27.8, 28.2, 28.7 & 189.228621& +62.238572 & 163.13\\
  GOODS-S & \citet{guo:2013} & 5 & 0.17 & 27.4, 28.2, 29.7 & 53.122751 & -27.805089 & 159.36\\
  UDS & \citet{galametz:2013} & 1 & 1 & 27.9 & 34.406250 & -5.2000000 & 195.58\\
  \hline
\end{tabular}
\caption{References and image characteristics for the published papers on the five observed CANDELS fields. The $\sigma$ column indicates whether limiting magnitudes were computed at $5\sigma$ or $1 \sigma$, and aperture provides the aperture used to compute the limiting magnitude (see Equation 5). Depths are for the F160W image. The quoted effective areas are for the ``wide" images. }
\label{tab:candelsfields}
\end{center}
\end{table*}

\subsection{Empirical model: {\sc UniverseMachine}}
The {\sc UniverseMachine} is an empirical model that connects galaxies' star formation rates to their host haloes' masses ($M_h$), accretion rates ($\dot{M}_h$), and redshifts \citep{Behroozi:2019}.  Using an initial guess for the distribution of galaxy SFRs as a function of host halo properties (i.e., $P(SFR|M_h,\dot{M}_h, z)$), it populates all haloes in a dark matter simulation with SFRs.  These SFRs are then integrated along merger trees to obtain galaxy stellar masses and luminosities.  The statistics of the resulting mock universe are compared to those from observations, including stellar mass functions ($z=0-4$), quenched fractions ($z=0-4$), cosmic SFRs ($z=0-9$), specific SFRs ($z=0-8$), UV luminosity functions ($z=4-10$), UV--stellar mass relations ($z=4-8$), auto- and cross-correlation functions for star-forming and quiescent galaxies ($z=0-1$), and quenched fractions of isolated galaxies as a function of environment ($z=0$).  Comparing these observables results in a likelihood for the guess for $P(SFR|M_h,\dot{M}_h, z)$.  This likelihood is given to a Monte Carlo Markov Chain algorithm to generate a new guess, and the process is repeated millions of times to obtain the posterior distribution of galaxy--halo connections that are consistent with all input observations.

The {\sc UniverseMachine} attempts to forward-model to available observations as much as possible.  This includes accounting for random and systematic errors in both stellar masses and SFRs, which can both rise to levels of $\sim 0.3$ dex even at intermediate redshifts ($1<z<3$).  As with other models in this paper, the {\sc UniverseMachine} uses an orphan prescription to extend the lifetime of infalling satellites.  Specifically, satellite lifetimes are extended until (or truncated after) their circular velocities reach $\sim 0.5$ of the value reached at peak mass; this ratio is calibrated to match $z=0-1$ galaxy autocorrelation functions.

\section{Mock Catalogs: Contents, documentation and access}
\label{sec:mockproperties}
The mock catalogs contain a large number of ``intrinsic'' physical
parameters, such as halo mass, stellar mass, star formation rate, etc,
as well as observable parameters such as magnitudes in filter bands
relevant to the CANDELS survey. A full list of the quantities
contained in the mock catalogs, with their units, is given in the
supplementary information that may be downloaded at \url{https://users.flatironinstitute.org/~rsomerville/Data_Release/CANDELS/mocklc_pub1.0.pdf}. 

A very large number of different instruments have been used to observe
the CANDELS fields. In many cases, different filters have been used
that are, in practice, very similar to one another. In order to
prevent our already very large catalogs from becoming even more
unwieldy, we have recorded only a representative set of filter
response functions. Typically, the slightly different filters used in the different CANDELS fields differ from the corresponding representative filter by less than 0.01 magnitude for all galaxies in our mock catalogs. The full set of filter response functions may be downloaded from \url{https://users.flatironinstitute.org/~rsomerville/Data_Release/CANDELS/filters/CANDELS.filters.tar}.

The above documentation and supplementary information, as well as access to flat files containing the Santa Cruz SAM mock lightcone files for all fields and all realizations is available at  \url{https://www.simonsfoundation.org/candels-survey}. All three sets of mock catalogs (SC-SAM, Lu-SAM, and UniverseMachine) may also be accessed through the Flatiron Institute Data Exploration and Comparison Hub (Flathub), at \url{http://flathub.flatironinstitute.org/group/candels}. Flathub allows the user to interactively select a subset of models, fields, realizations, and/or catalog columns for download, and to pre-filter the data before downloading. This is a convenient option for users who are only interested in a subset of the models, or specific quantities or types of objects, since the total data volume is quite large. 
\begin{figure*} 
\begin{center}
\includegraphics[width=\textwidth]{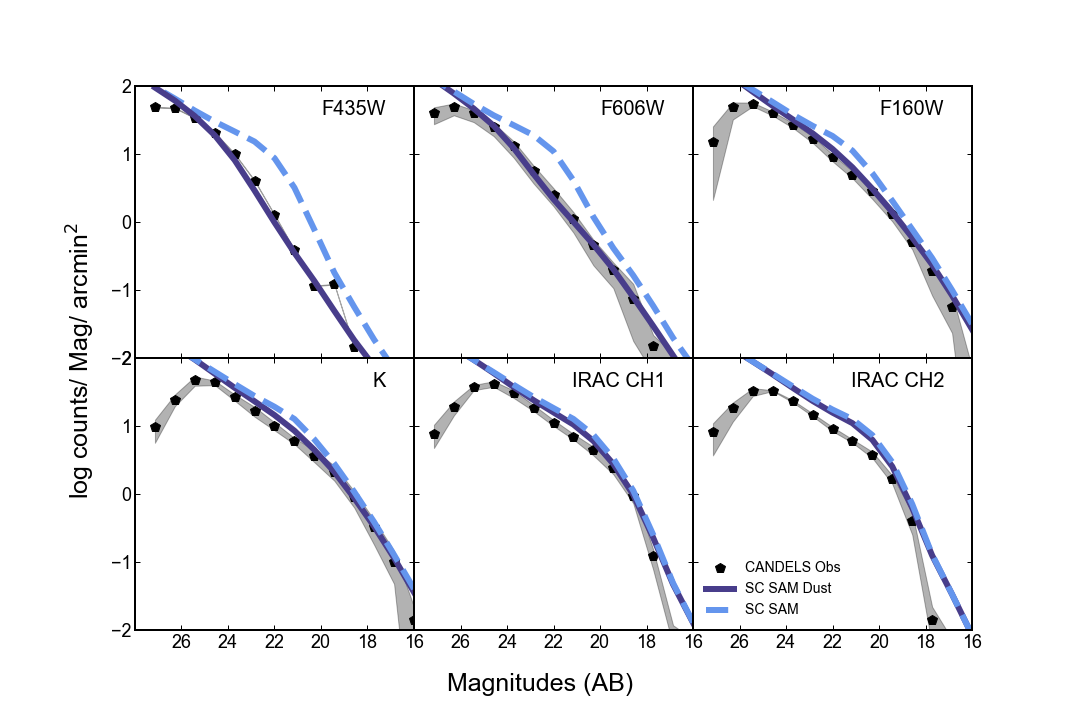}
\end{center}
\caption{Total counts as a function of apparent magnitude in different observed frame filters as indicated by the labels in each panel, for the SC SAM mock catalogs compared with CANDELS. Solid dark blue lines show the SC SAM predictions with dust attenuation included, and dashed light blue lines show the SC SAM predictions without dust. The grey shaded region shows the range of values between the different CANDELS fields, and the black symbols show the median of the values in all four fields. No magnitude cuts or completeness corrections have been applied to either the models or observations, and the observed counts become incomplete at around magnitude 25.5 or 26. The SC SAM predictions match the observations well in the F435W, F606W, and F160W bands, but a bit less well in the redder K and IRAC bands.
\label{fig:counts}}
\end{figure*}

\section{CANDELS observations and Theory Friendly Catalogs}
\label{sec:tfcandels}

The CANDELS survey is a 902-orbit legacy program which carried out imaging with the WFC3 camera on HST in five fields: COSMOS, EGS, GOODS-N, GOODS-S, and the UDS, over a combined area of about 0.22 deg$^2$. Each field has a different suite of ancillary imaging data from X-ray to radio from the ground and space, which have been incorporated into multi-wavelength catalogs and used to estimate photometric redshifts and physical properties such as stellar masses and star formation rates. Please see \citet{grogin:2011} and \citet{koekemoer:2011} for details of the survey design and basic image processing, and the five ``field'' papers (summarized in Table~\ref{tab:candelsfields}) for details on the catalog construction for each field. The CANDELS catalogs released by the team may be accessed at \url{https://archive.stsci.edu/prepds/candels/}, and an interactive web-based portal to some of the CANDELS catalog and image data is available at \url{https://rainbowx.fis.ucm.es/Rainbow_Database/Home.html}. 

We have created a curated version of the CANDELS high level science products, which have been designed to be easy to use for comparisons with theoretical models and simulations. The format and contents of the CANDELS ``theory friendly catalogs'' (TF-CANDELS) has been standardized and homogenized over all five fields, and the catalogs have had a standard set of flags and cuts applied. Each theory friendly catalog contains a standardized set of observed frame and rest-frame photometry, along with redshifts, structural parameters (size and Sersic index), and multiple stellar mass and star formation rate estimates. In the original catalogs, the ``value added'' quantities such as photometric redshifts, structural parameters, and stellar masses are all in separate files which must be joined. 

The photometric redshifts in the TF catalogs are the updated estimates from Kodra et al. (in prep), and these are used for all derived quantities in the TF-CANDELS catalogs that depend on redshift (e.g. absolute magnitudes, stellar masses, SFR). We have checked, however, that none of the results shown in this paper differ significantly from those that are obtained using the published team redshifts as documented in \citet{dahlen:2013}. 

Rest-frame absolute magnitudes were computed for the same filter response functions used to compute rest frame photometry in the mock catalogs. Rest-frame magnitudes were computed using the package EAZY \citep{EAZY}, with the details of the set-up and parameter file as specified in the TFCD (Appendix E). In addition, the TF-CANDELS catalogs provide alternate estimates of the absolute magnitudes computed using the {\sc zphot} package\footnote{Note that unlike all other magnitudes in the catalogs, these are in the Vega system.} \citep{Fontana:2000,Merlin:2019}, and of $U-V$ and $V-J$ colors computed using the SED-fitting method of \citet[][hereafter P12]{Pacifici:2012}. The TF-CANDELS catalogs also include stellar masses estimated using both the {\sc zphot} and P12 approaches. For star formation rates (SFR), in addition to estimates based on {\sc zphot} and P12, the catalogs also include the SFR estimates presented by \citet{barro:2019}, which utilize either a combination of rest-UV and mid-IR photometry, for galaxies that are detected in the IR, or a dust-corrected estimate based on the rest-UV. A detailed comparison of how these derived quantities differ for the different methods is not in the scope of this paper, however, we do comment briefly on this issue in the discussion (Section~\ref{sec:discussion}). 

The original files and catalog field names used to create each entry in the TF-CANDELS catalogs are specified in the TF-CANDELS Documentation (TFCD Appendix A; \url{https://users.flatironinstitute.org/~rsomerville/Data_Release/CANDELS/TFCD.pdf}). We have selected a ``representative'' observed U-band and K-band filter for each field. The fields were observed with different telescopes and different instruments, so in practice the actual filters differ a bit from field to field. The details of the actual filters used for each field are provided in Appendix B of the TFCD.

Appendix C of the TFCD describes how we carried out the calculation of F160W limiting magnitude for each object in the TF-CANDELS catalogs. Using the F160W weight maps for each field,  we computed the average RMS as $\langle RMS \rangle = \sqrt{(1/\langle w_i \rangle}$, where $\langle w_i \rangle$ is the average weight over a 6x6 square of pixels surrounding the center of each galaxy. We then computed the limiting magnitude as
\begin{equation}
m_{\rm lim} = -2.5 \log_{10} (\sqrt{A_1 \langle RMS^2 \rangle} + z_p 
\end{equation}
where $A_1 = 1/(0.06 {\rm arcsec/pixel})^2$, and the zeropoint $z_p$ is given by
\begin{multline*}
z_p = -2.5 \log_{10}(PHOTFLAM) \\ 
-5\log_{10}(PHOTPLAM) - 2.408
\end{multline*}
PHOTFLAM is the conversion factor from counts/s to erg s$^{-1}$ cm$^{-2}$ $\AA^{-1}$, and PHOTPLAM converts from flux per unit wavelength $f_\lambda$ to flux per unit frequency $f_\nu$. The values are taken from the image headers. In the TF catalogs, the limiting magnitude is defined as the 1$\sigma$ limiting magnitude within an aperture with area 1 arcsec$^2$. Some of the CANDELS field papers have used other definitions, such as the 5$\sigma$ limiting magnitude within a different aperture (see Table~\ref{tab:candelsfields}). In order to compute the limiting magnitude at 5$\sigma$ within some other aperture, one can adopt:
\begin{equation}
  m_{\rm lim} = -2.5 \log_{10} (\sqrt{5 A \langle RMS^2 \rangle} + z_p
\end{equation}
where $A$ is the area of the desired aperture in pixels. 

Cuts were applied to remove objects with flags indicating bad photometry, and the field CLASS\_STAR was used to remove likely stars (see TFCD Appendix D).

We computed effective areas for various F160W limiting magnitude cuts in the following manner (see also Appendix F of the TFCD). The CANDELS weight maps were transformed into limiting magnitude maps using the formulae given above, and the effective area was calculated for each limiting magnitude bin by adding up the number of pixels for which the limiting magnitude was at the bin's value or fainter. Tables of these calculated effective areas (in arcmin) vs. the F160W limiting magnitude are provided at \url{https://users.flatironinstitute.org/~rsomerville/Data_Release/CANDELS/effarea/}.

All of the above documentation and supplementary information, as well as access to flat files containing the theory friendly catalogs for all five fields are linked from the landing page \url{https://www.simonsfoundation.org/candels-survey}. In addition, the TF-CANDELS catalogs can be previewed and downloaded through  \url{http://flathub.flatironinstitute.org/group/candels}.

\section{Results}
\label{sec:results}

\begin{figure*} 
\begin{center}
\includegraphics[width=\textwidth]{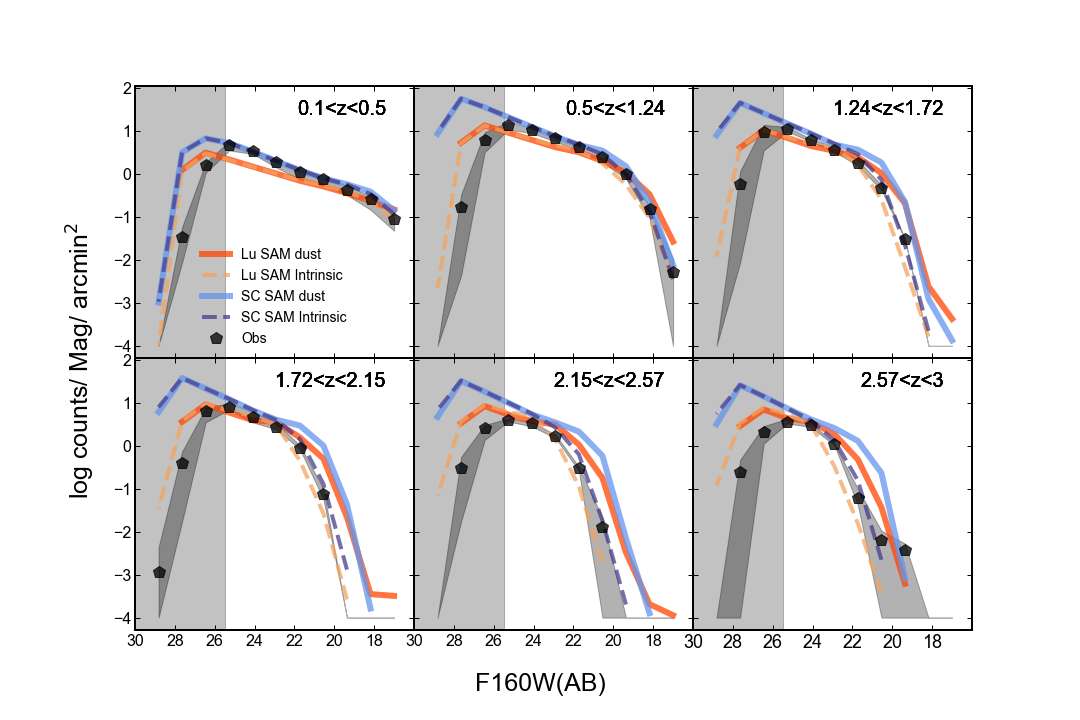}
\end{center}
\caption{Counts as a function of apparent observed frame F160W magnitude split into redshift bins, for the Lu (orange) and SC (blue) SAM mocks compared with CANDELS (black symbols show the mean over all four fields; dark gray shaded areas show the minimum and maximum value in each bin over the four fields). Light gray vertical shaded regions show approximately where the CANDELS Wide observations are expected to be incomplete. Dashed lines show the intrinsic counts before dust attenuation has been added to the model galaxies, and solid lines show the predictions including dust attenuation. The agreement between the predicted and observed counts is qualitatively good in all redshift bins. 
\label{fig:counts_redshift}}
\end{figure*}

In this section we compare the predictions of the mock catalogs with CANDELS observations in different redshift bins. We investigate quantities in `observational' space such as counts in the observed H$_{160}$ (F160W) band as well as derived quantities such as rest-frame luminosity functions and color-color diagrams. We further investigate comparisons with physical properties derived from SED fitting to the CANDELS data, such as stellar mass functions and SFR functions. For this analysis, we adopt a standard set of redshift bins: $0.1$, $0.5$, $1.24$, $1.72$, $2.15$, $2.57$, $3.0$. These have been chosen so that all except for the lowest redshift bin have roughly equal comoving volume.
In all of the results presented below, the model results are obtained by averaging over all eight realizations of all five fields (40 lightcones in all, covering a total area of 40 times 5362 arcmin$^2$, or $\simeq 60$ sq. deg.). 

The observational results shown in all figures to follow are obtained by averaging over the EGS, GOODS-S, GOODS-N, and UDS fields, and the shaded areas show the minimum and maximum values of the binned quantity in each bin from field to field. We omitted the COSMOS field from this analysis because we found that the counts in redshift bins $z\gtrsim 1.5$ were anomolously low compared with the other four fields (we are in the process of investigating the reason for this discrepancy). The TF catalogs were cut at a limiting magnitude of 27.6, and only objects with  F160W$<25.5$ are plotted, unless noted otherwise. Areas were computed using the effective area tables for the corresponding depth provided in the supplementary materials. We adopt the P12 estimates for the stellar mass and SFR as defaults. 

\subsection{Observed-frame counts as a function of apparent magnitude}

Figure~\ref{fig:counts} shows the counts (number of objects per bin in apparent magnitude, per sq. arcminute on the sky) integrated over all redshifts, for a selection of the CANDELS observed frame filter bands. As the Lu mocks do not include IRAC photometry, we only show the comparison with the SC SAMs here --- the Lu SAMs produce similar results in the F435W, F606W, F160W, and K bands. No magnitude cuts or completeness corrections have been applied to either the models or observations, and the observed counts become incomplete at around magnitude 25.5 or 26. The SC SAM predictions match the observations well in the F435W, F606W, and F160W bands, but less well in the redder K and IRAC bands. This is in part because there is flexibility to match the bluer bands by adjusting the dust correction. It may also reflect the presence of an older stellar population in the model galaxies than is present in the real Universe, as we will discuss further below. This figure illustrates the potential to calibrate models using multiband photometry instead of derived quantities such as stellar masses. 

Figure~\ref{fig:counts_redshift} shows the counts in the HST F160W filter (which is the detection filter for CANDELS) split into redshift bins. The model results are shown for the intrinsic fluxes without accounting for attenuation by dust, and including the model for dust attenuation described in Section~\ref{sec:stellpop}. Overall, there is very good agreement between the models and observations once dust attenuation is accounted for, for magnitudes where the CANDELS wide catalogs are highly complete (F160W $< 25.5$). For fainter magnitudes, the CANDELS catalogs are incomplete and the photometric redshifts also become unreliable. The turnover in the model curves at faint magnitudes is due to the limited mass resolution of the N-body simulation on which the SAM catalogs are built. As noted in Section~\ref{sec:sams}, the SC SAMs effectively have higher resolution than the Lu-SAMs, as they make use of EPS-based merger trees rather than the N-body based trees, which is the main reason that the counts turn over at a slightly fainter magnitude. 

\subsection{Rest-frame Luminosity Functions}

\begin{figure*} 
\begin{center}
\includegraphics[width=\textwidth]{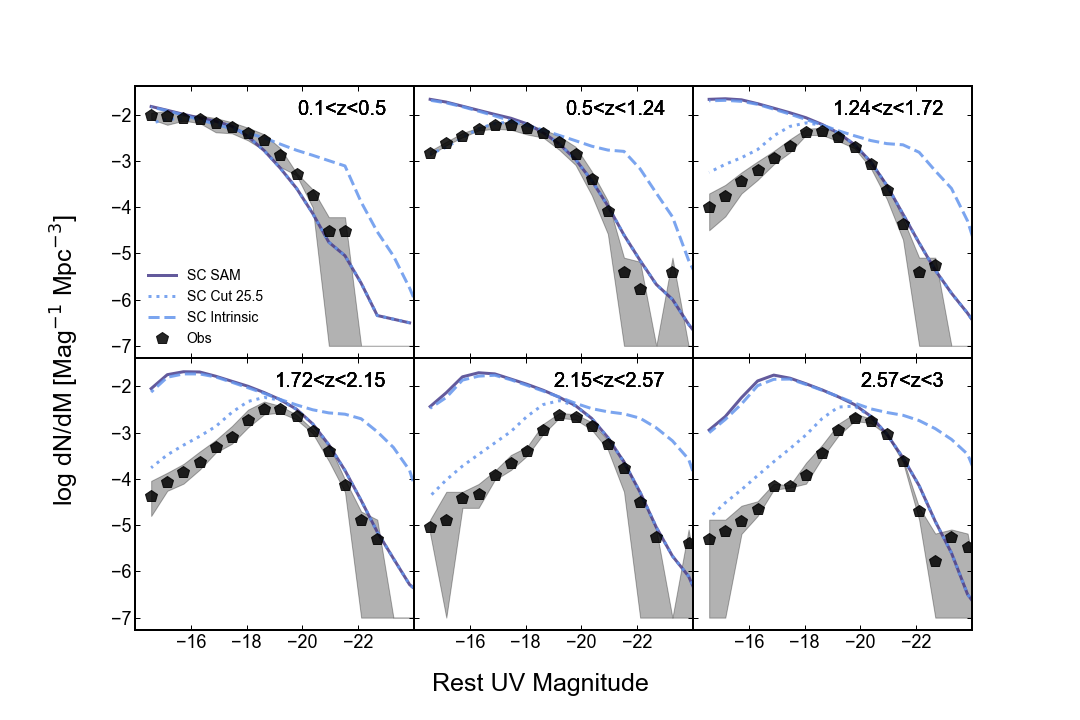}
\end{center}
\caption{Luminosity functions in the rest-frame UV ($1500$ \AA) divided into redshift bins, for
  the SC SAM, compared with the corresponding distributions from CANDELS (black symbols show the mean over all fields; shaded areas show the minimum and maximum value in each bin over the four fields). Dashed lines show the intrinsic luminosity functions with no dust attenuation; solid lines show the model predictions with dust attenuation included. Dotted lines show the dust attenuated model predictions with a cut of F160W$<25.5$ applied (similar to the observations). The SC SAM predictions agree with the observed distributions quite well in the regime where the observations are highly complete. 
\label{fig:lf_UV_SC}}
\end{figure*}

Figure~\ref{fig:lf_UV_SC} shows the binned histograms of rest-frame absolute magnitude in the rest-$1500$ \AA band, for the CANDELS observations (using the EAZY-based estimate) and the SC SAM, in our standard redshift bins. We note that although we refer to these as ``luminosity functions", we have made no attempt to correct these for incompleteness as is generally done in the literature. We simply apply a cut of F160W$<25.5$ and bin in absolute luminosity. For the models, we show results both without and with modeling of dust attenuation. In addition, we show the model predictions both for the full sample and with a cut of observed frame F160W$<25.5$ (including dust), similar to the cut applied to the observational sample. We remind the reader that the CANDELS Wide sample is expected to be highly complete at this magnitude limit.  Similar comments apply to the V-band ``luminosity functions". Figure~\ref{fig:lf_UV_Lu} shows the same comparison for the Lu SAMs. The SC SAMs show very good agreement with the bright end of the UV LF when dust attenuation modeling is included; and the models with an F160W$<25.5$ cut show good agreement with the turnover and faint end of the UV LF as well. The Lu SAMs also show good agreement, but slightly underpredict the observed number density of faint galaxies in the lowest redshift bins (see Fig. \ref{fig:lf_UV_Lu}). 

\begin{figure*} 
\begin{center}
\includegraphics[width=\textwidth]{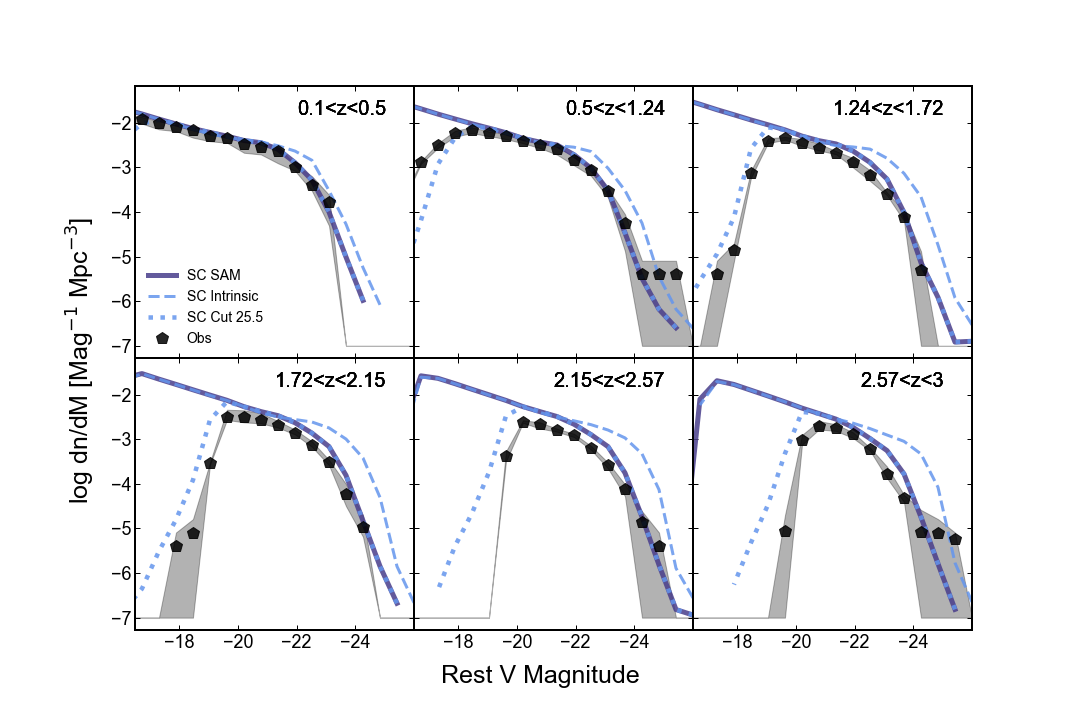}
\end{center}
\caption{Luminosity functions in the rest-frame V-band divided into
  redshift bins for the SC SAM, compared with the corresponding distributions from CANDELS. Key is as in Fig.~\ref{fig:lf_UV_SC}. The SC SAM predictions agree with the observed rest-frame V-band magnitude distributions fairly well in the regime where the observations are highly complete. 
\label{fig:lf_V_SC}}
\end{figure*}

Figure~\ref{fig:lf_V_SC} and \ref{fig:lf_V_Lu} show the rest-V band luminosity functions for the CANDELS observations (using the EAZY-based estimate) and the SC and Lu SAM models, with the same permutations as before. Once again, both models are in very good agreement with the observations once dust attenuation and a cut on $F160W$ are applied. In the SC SAMs, there is a small but significant excess of faint galaxies (fainter than $L_*$) relative to the CANDELS observations at redshifts $z\gtrsim 1.5$, but the agreement on the bright end is very good. The Lu SAMs show better agreement for faint galaxies, but show a possible deficit of bright galaxies at $z\gtrsim2$ -- but the model predictions are mostly within the range of values seen across the four different fields (see Fig. \ref{fig:lf_V_Lu}). The constraints on the models could be improved by adding observations from larger-area surveys. 

\subsubsection{Color-magnitude and color-color relations}

\begin{figure*} 
\begin{center}
\includegraphics[width=\textwidth]{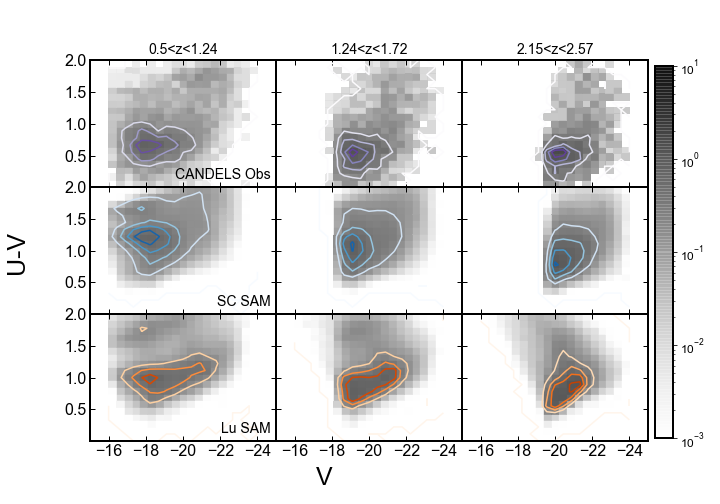}
\end{center}
\caption{The greyscale and overlaid contours show joint distributions of rest-frame $U-V$ color versus
  rest $V$ magnitude for the CANDELS observations (top) and the SC (middle) and Lu (bottom) SAMs, in three redshift bins. Both the observed and model galaxies are selected to have F160W$<25.5$, where the CANDELS Wide samples are highly complete. Both SAMs predict colors for low-luminosity galaxies that are up to $\sim 0.5$ dex redder than the observed colors, with an increasing discrepancy towards lower redshift.  
\label{fig:UVV}}
\end{figure*}

For both results presented in this sub-section, a magnitude cut of F160W$<25.5$ has been applied to both the model and observational samples, and dust attenuation is included in the model magnitudes. The CANDELS Wide samples used in this study should be highly complete at this magnitude limit. We use the P12 estimates from CANDELS for both $U-V$ and $V-J$ colors, and the EAZY estimate for the rest-frame V-band magnitude.

Figure~\ref{fig:UVV} shows the distribution of the number density of galaxies in the rest-$V$ magnitude versus $U-V$ color plane, in three redshift bins.  The overall distributions in color-magnitude space appear similar between the models and observations, although some quantitative differences are apparent. The main population of galaxies in the models is shifted to the red relative to the observations, by as much as 0.5 dex. The discrepancy increases towards lower redshifts.  The colors in the SC SAM are slightly redder than in the Lu SAM. The Lu SAM does not show as pronounced a trend between luminosity and color as the observations or the SC SAM. Both the Lu SAM and, to a lesser extent, the SC SAM, show a bimodal distribution of colors at faint luminosities. The population with redder colors is associated with satellite galaxies, and reflects a well-known tendency of SAMs to produce over-quenched low-mass satellite galaxies. However, even when comparing only central galaxies with the observed colors, there is a significant discrepancy in the sense of the SAMs producing galaxies with colors that are too red. This is due to the known tendency in these SAMs for low-mass galaxies to form too many stars at high redshift (and so to have too large an old stellar population) while being too inefficient at forming stars at low redshift \citep{white:2015}. The effect of dust reddening on colors is also very uncertain.

\begin{figure*} 
\begin{center}
\includegraphics[width=\textwidth]{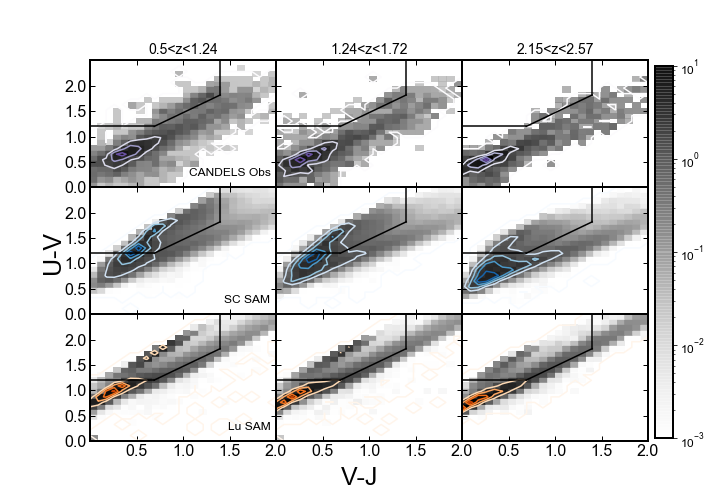}
\end{center}
\caption{Distribution of galaxies in the $U-V$ vs. $V-J$ plane for the
  CANDELS observations (top row), the SC SAM (middle row), and the Lu SAM (bottom row). Both the observed and model galaxies are selected to have F160W$<25.5$, where the CANDELS Wide samples are highly complete. The solid black line shows the region of this diagram that is typically associated with quiescent galaxies \protect\citep{Williams:2009}. 
\label{fig:UVJ}}
\end{figure*}

Figure~\ref{fig:UVJ} shows the distribution in rest-frame $U-V$ versus $V-J$ color space for the CANDELS observational sample and the SC SAM and Lu SAM, in three redshift bins. This diagram is often used to identify and separate star forming galaxies and quiescent galaxies, where quiescent galaxies are expected to be located in the upper left-hand region of the plot, and a nominal dividing line is shown. Once again we can clearly see that the star forming population in the models is too red in $U-V$, but is in better agreement in $V-J$ (which is less sensitive to stellar age). We also see that the standard dividing line between quiescent and star forming galaxies does not separate these populations very effectively in the models, perhaps reflecting shortcomings in the dust modeling, or differences in the ensemble of star formation histories. See \citet{Brennan:2015} for a detailed analysis of the quiescent fraction in the the Santa Cruz SAMs compared with CANDELS observations, and \citet{Pandya:2017} for an analysis of the transition galaxy population and the quenching rate in the SC SAMs and CANDELS.

\subsection{Stellar Mass functions, Star Formation Rate functions and the Star Formation Sequence}

\begin{figure*} 
\begin{center}
\includegraphics[width=\textwidth]{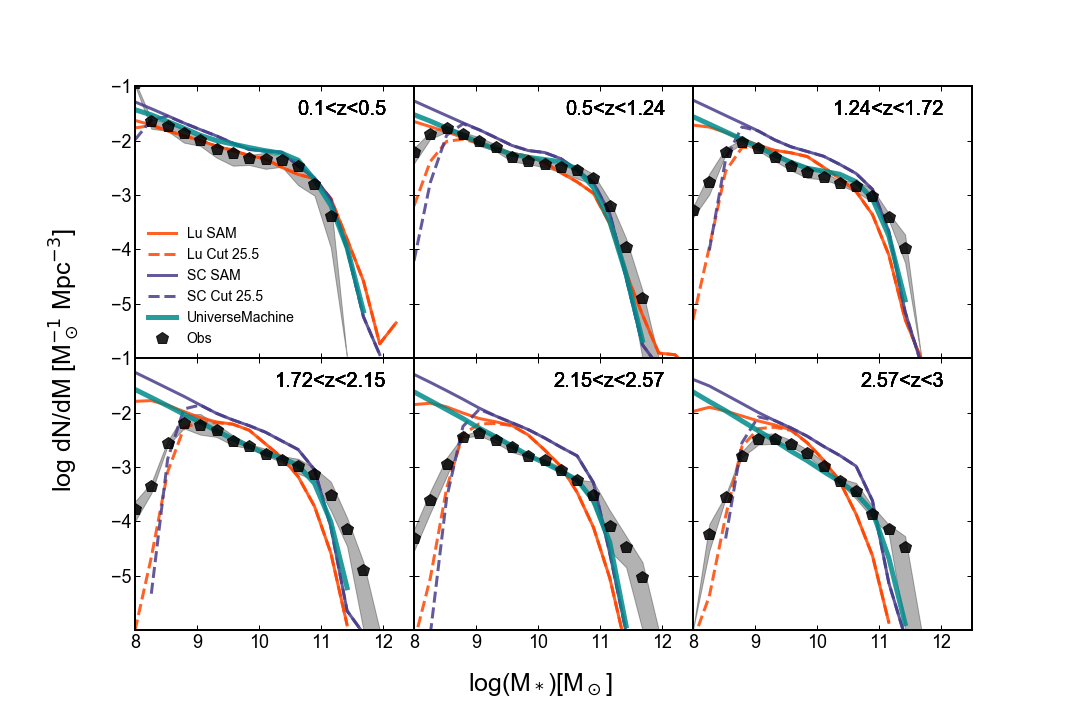}
\end{center}
\caption{Stellar mass functions divided into redshift bins, for
  the two SAMs and {\sc UniverseMachine}, compared with stellar mass distribution functions derived from CANDELS. The CANDELS observations are shown for F160W$<25.5$ and are not corrected for incompleteness. The SAM mass functions are shown for all galaxies (solid) and for galaxies with F160W$<25.5$ (dashed). The stellar masses in all models shown here do not account for observational errors. Both SAMs predict an excess of low-mass galaxies at intermediate redshift ($1 \lesssim z \lesssim 3$), and perhaps a deficit of massive galaxies at $z\gtrsim 1.5$. 
\label{fig:SMF}}
\end{figure*}

Figure~\ref{fig:SMF} shows stellar mass functions for the CANDELS observations, using stellar mass estimates based on the P12 method, compared with stellar mass functions from the SC SAM, Lu SAM, and {\sc UniverseMachine}. We note that {\sc UniverseMachine} is an empirical model that was calibrated to match previous estimates of the galaxy stellar mass function over a wide range of redshifts \citep[see][for details]{Behroozi:2019}, so this is an indirect way to compare the SAM predictions and the CANDELS measurements with previous estimates of the stellar mass function. It is important to note that the CANDELS observational ``stellar mass functions" have not been corrected for incompleteness and only galaxies with F160W$<25.5$ have been plotted. The SAM predictions are shown for all galaxies and also with a cut of F160W$<25.5$, as in the observations. The turnover in the stellar mass function occurs at a similar mass in the SAMs and in the CANDELS observations.  The {\sc UniverseMachine} mass functions are in good agreement with the CANDELS estimates, except at the high mass end. This is due to errors on the stellar masses, which cause an Eddington bias that make the high-mass end of the SMF shallower. 
This is illustrated in Figure~\ref{fig:SMF_obs}, which shows the {\sc UniverseMachine} predictions with stellar mass errors included as described in \citet{Behroozi:2019}. Here it can be seen that the stellar mass errors can have a significant effect on the high-mass end of the SMF, especially at high redshift. With the observational errors included, {\sc UniverseMachine} is in near-perfect agreement with the CANDELS stellar mass distributions, as expected. 
However, estimating the error in stellar mass and how it depends on other galaxy properties in detail is highly non-trivial, so we do not include errors on the stellar masses in the SAM mock catalogs. The SC SAMs agree well with {\sc UniverseMachine} intrinsic stellar masses at the high mass end, but systematically overproduce low-mass galaxies (below the knee in the SMF) at all redshifts, but to an increasing degree at high redshift. This is a well-known and widespread problem with many current models of galaxy formation, which is caused by too-early formation of stars in low-mass galaxies \cite[see][for a detailed discussion]{somerville:2015,white:2015}. The Lu SAM shows better agreement with the abundance of low-mass galaxies, but still overproduces them in the highest redshift bins, and may underproduce massive galaxies at high redshift. However, it is impossible to rigorously assess the agreement at the high-mass end for both SAMs due to the uncertainty in the errors on the observational estimates of the stellar masses, as well as uncertainties due to field-to-field variance, which are only crudely indicated here.

\begin{figure*} 
\begin{center}
\includegraphics[width=\textwidth]{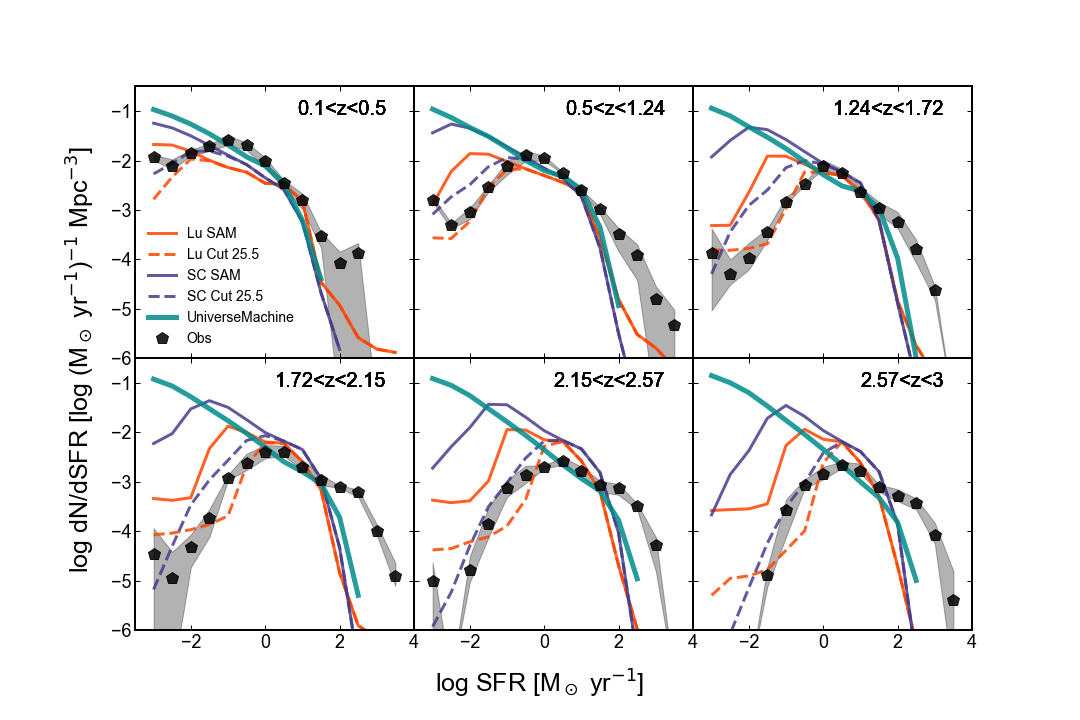}
\end{center}
\caption{Star formation rate distribution functions divided into redshift bins, for
  the two SAMs and {\sc UniverseMachine}, compared with SFR functions derived from CANDELS. The CANDELS observations are shown for F160W$<25.5$ and are not corrected for incompleteness. The SAM SFR distribution functions are shown for all galaxies (solid) and for galaxies with F160W$<25.5$ (dashed). The model SFR predictions do not account for observational errors. The CANDELS SFR distribution has a large excess of high-SFR galaxies relative to both SAMs and {\sc UniverseMachine} predictions. 
\label{fig:SFRF}}
\end{figure*}

Figure~\ref{fig:SFRF} shows star formation rate distribution functions for the CANDELS observations, using SFR estimates based on the method of P12, compared with SFR functions from the SC SAM, Lu SAM, and {\sc UniverseMachine} (intrinsic values of SFR, without observational errors, are shown). 
The SC and Lu SAM SFR have been averaged over a timescale of 100 Myr, while the P12 SFR estimates are averaged over 10 Myr; however, we do not expect this to cause large differences.
CANDELS observational SFR functions have not been corrected for incompleteness and only galaxies with F160W$<25.5$ have been plotted. The SAM predictions are shown for all galaxies and also with a cut of F160W$<25.5$, as in the observations. The amplitude and location of the ``knee" of the SFR distribution function agrees well between both SAMs, {\sc UniverseMachine}, and the observations. The predictions of both SAMs and {\sc UniverseMachine} are very similar for the high SFR part of the distribution. At $z\gtrsim 1.7$, both SAMs predict a higher amplitude and steeper distribution below the knee than UM and the CANDELS observations. However, the SFR distribution derived from CANDELS is significantly higher in amplitude above SFR values of $\sim 100$ \msun yr$^{-1}$ than any of the model predictions, by as much as several orders of magnitude. Fig.~\ref{fig:SFRF_obs} again shows the SFR distribution for CANDELS and for {\sc UniverseMachine} predictions with and without observational errors added. Again, the observational errors cause a small increase in the amplitude at the high SFR end, but based on the assumed magnitude of the errors on SFR from \citet{Behroozi:2019}, this cannot fully account for the discrepancy between the models and observations. 

\begin{figure*} 
\begin{center}
\includegraphics[width=\textwidth]{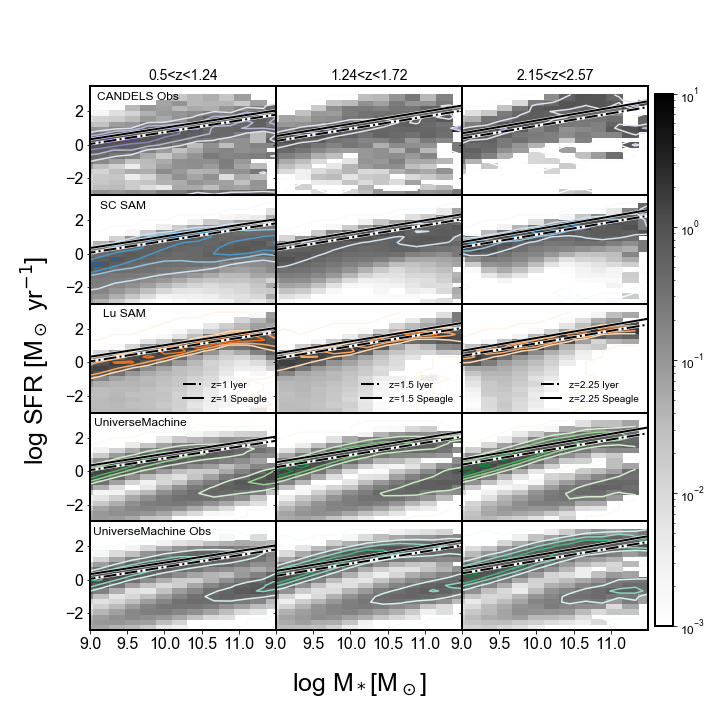}
\end{center}
\caption{Greyscale and overlaid contours show the conditional distribution of star formation rate for a given stellar mass, for CANDELS observations (top row), the two SAMs (SC; second row, and Lu; third rows), and {\sc UniverseMachine} with and without observational errors included (bottom two rows). Relations from the literature \citep{speagle:2014,Iyer:2018} are overplotted. In both SAMs, the predicted SFR are too low at fixed stellar mass, particularly for low-mass galaxies at $z \lesssim 1$.
\label{fig:SFMS}}
\end{figure*}

Fig.~\ref{fig:SFMS} shows the conditional distribution of SFR for a given stellar mass  in several redshift bins, for CANDELS using stellar mass and SFR estimates from P12, and for both SAMs and for {\sc UniverseMachine} without and with observational errors included. SFR-stellar mass sequence relations from the literature \citep{speagle:2014,Iyer:2018} are also overplotted. In the first two redshift bins shown, the CANDELS results are consistent with the literature sequence, while in the highest redshift bin, there is a population of massive galaxies that lies above the literature sequence. The median SFR at a given stellar mass is systematically lower than the literature relations and CANDELS in both SAMs, more so in the SC SAM, and has a steeper slope, such that low-mass galaxies lie below the observed SFR sequence. This is a further reflection of the same problem that caused the colors in the SAMs to be overly red, namely, SFR is too strongly suppressed in low-mass galaxies in the SAMs. 
Interestingly, {\sc UniverseMachine} also shows a mild steepening in the SFR sequence at low-mass, but to a much lesser extent than the SAMs. Observational estimates of complete samples of such low-mass galaxies are extremely challenging to obtain, but these results suggest it may be interesting to do so.  

\begin{figure*} 
\begin{center}
\includegraphics[width=\textwidth]{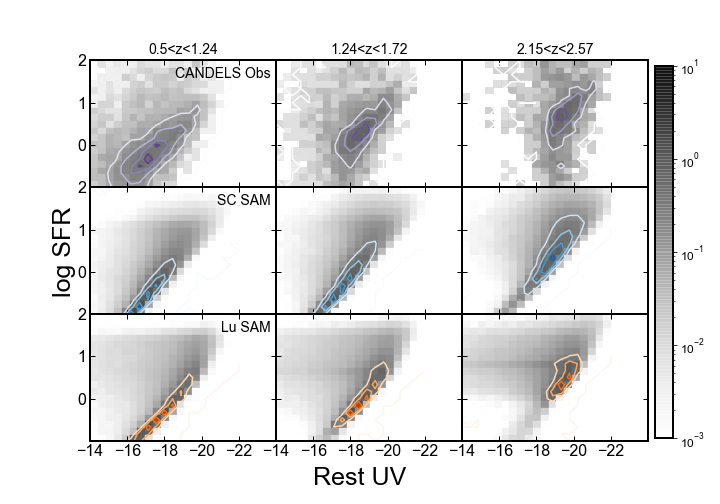}
\end{center}
\caption{Greyscale and overlaid contours show the distribution of log SFR vs. rest-frame UV magnitude, for CANDELS observations (top row) and the two SAMs (SC; second row, and Lu; third row). The CANDELS SFR estimates at a given rest-UV magnitude are significantly higher than the predictions of the SAMs.
\label{fig:SFR_restUV}}
\end{figure*}

In order to try to interpret the very different conclusions we might reach from comparing the rest-UV luminosity function of galaxies (which shows excellent agreement between model predictions and observations) and the SFR function (which shows disagreement between the models predictions and observations at the level of multiple orders of magnitude), we examine the relationship between rest-UV luminosity and SFR in the CANDELS observations (using the P12 estimates of SFR) and in the semi-analytic models (where the rest-UV magnitude includes dust). Figure~\ref{fig:SFR_restUV} shows this relationship, and reveals that the P12 SFR estimates in CANDELS are significantly higher for a given rest-UV magnitude than the predictions of the SAMs, especially at high redshift. This helps to reconcile the different conclusions that we might reach from comparing the observed and predicted rest UV luminosity functions and SFR functions, but begs the question as to the reason the relationship between rest-UV magnitude and SFR is so different. The main possibilities are the assumed/estimated dust attenuation and the star formation history. We investigate the former possibility in the next sub-section. 

\subsection{Dust Attenuation}

\begin{figure*} 
\begin{center}
\includegraphics[width=\textwidth]{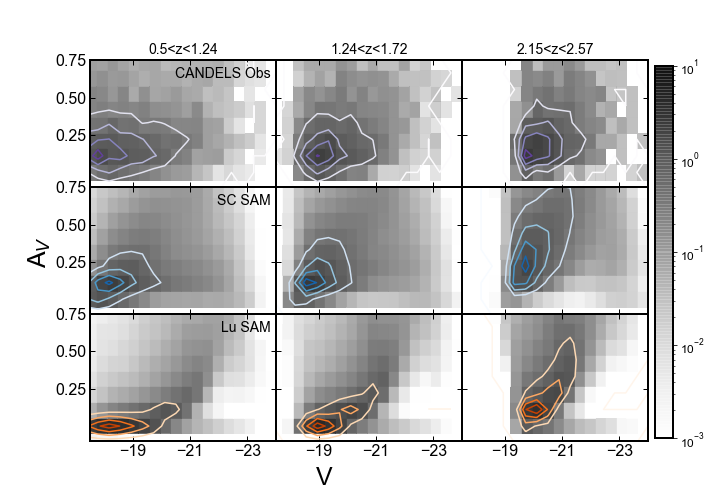}
\end{center}
\caption{The greyscale and overlaid contours show the joint distribution of dust attenuation in the rest V-band and rest-V band magnitude, in different redshift bins, as estimated from SED fitting using the method of P12 (top row), and as added to the model galaxies in the SC (middle) and Lu (bottom row) SAMs. It is encouraging that the dust attenuation adopted in the SAMs is similar to the results from SED fitting. 
\label{fig:AVV}}
\end{figure*}

Dust attenuation is an important ingredient in forward modeling the semi-analytic models to the observational plane. By the same token, it is a critical ingredient in SED fitting methods used to estimate physical properties from galaxy photometry. As described in Section~\ref{sec:stellpop}, the normalization of the relationship between dust optical depth in the V-band and galaxy properties such as gas surface density and metallicity in the SAMs has been adjusted empirically to match the observed UV, B, and V-band luminosity functions. Therefore it is interesting to see how this quantity compares with the dust attenuation derived from SED fitting to the CANDELS observations using the method of P12. Figure~\ref{fig:AVV} shows the attenuation in the rest V-band versus the (attenuated) V-band, in three redshift bins, for the SED-fitting derived results from CANDELS and for the SC and Lu SAMs. The medians of the distributions are quite similar. In the two higher redshift bins, the SAMs show a stronger trend between V-band magnitude and attenuation than the CANDELS estimates.  Both the SAMs assume a fixed dust attenuation curve, while the SED fitting procedure of P12 adopts a two component dust model.

\section{Summary and Discussion}
\label{sec:discussion}

The mock lightcones that we have presented here enable a more detailed comparison with observations than has often been done in the past. We find extremely good agreement between the observed frame counts predicted by the SAM and observations for the F435W, F606W, F160W, and K bands, and less precise but still good ($\sim$ 0.3 dex or better) for the counts in the IRAC Ch1 and Ch2 bands. The agreement with observed frame F160W counts split by redshift is excellent (better than 0.1 dex for the SC SAM where the observations are highly complete). Agreement between predicted and observed rest-frame UV and V-band luminosity functions from $0.1 < z < 3$ is also very good, everywhere better than $\sim$0.3 dex. Agreement between predicted and observed rest-frame U-V colors is less successful; the SAM predictions are as much as $\sim 0.5$ too red for low-luminosity galaxies at low redshift $z\lesssim 1$. The agreement between the SAM predicted SMF and that derived from CANDELS via SED fitting is good but notably poorer quantitatively than for the luminosity function comparison. Even more dramatically, the SAM predictions for the SFR function show very large discrepancies with the CANDELS SFR functions derived from SED fitting using the method of P12 (up to several orders of magnitude), presenting a very different picture from that obtained through comparing the rest-UV luminosity functions, which are in excellent agreement as noted above. We show that, as expected, the SFR-$L_{\rm UV}$ relationship predicted by the SAM and derived in CANDELS via SED fitting are very different, likely reflecting either different assumptions about dust attenuation and/or the galaxy star formation histories. 
 
Although semi-analytic models of galaxy formation are known to reproduce many key observations, the current generation of models is also known to show some tensions with observations that have been discussed extensively in the literature. These tensions include 1) models tend to \emph{overproduce} low-mass galaxies at intermediate redshifts $1 \lesssim z \lesssim 2$ relative to observations; 2) models tend to \emph{underproduce} massive galaxies at $z\gtrsim 1$; 3) model galaxies at $1\lesssim z \lesssim 4$ have SFR that tend to be \emph{lower} than observational estimates. It can be seen in the compilation presented in figure~4 and 5 of \citet{SD15} that these discrepancies are common not only to most semi-analytic models but also to several large-volume hydrodynamic simulations. So far, these problems have been overcome only by explicitly tuning the models to match observational constraints at high redshift, as in \citet{Henriques:2015}. Taken together, these discrepancies suggest a picture in which star formation is \emph{not efficient enough} in massive galaxies at intermediate redshifts ($1\lesssim z \lesssim 4$), and is \emph{too efficient} in low-mass galaxies in this same redshift range. At the same time, star formation rates are \emph{too low} in low mass galaxies at low redshift ($z\lesssim 1$). This leads to low-mass galaxies with too large an old stellar population (and too few young stars), which is likely the reason that the predicted colors in the models are too red for low-luminosity galaxies. 

There of course remain many uncertainties in the physical processes that are implemented in these models, as discussed extensively in recent reviews on the subject of physics-based cosmological models of galaxy formation \citep{SD15,naab:2017}. Some of the most important processes include the efficiency with which stellar driven winds can transport mass, energy, momentum, and metals out of the ISM and into the CGM and IGM, and the timescale on which ejected material returns to galactic halos or to the ISM. The treatment of these processes in cosmological simulations and SAMs has been highly phenomenological up until now. But recently, simulations in which the Sedov-Taylor blastwaves from supernovae explosions are explicitly resolved (or semi-resolved) have been used to extract the emergent wind launching characteristics \citep[][Pandya et al. in prep]{kim:2020a,kim:2020b}. Much progress can be made by using these wind scalings as input into larger volume cosmological simulations, and we can hope that this will help to resolve some of the tension with observations seen in this study and others. Other key areas where development effort is needed on the theoretical modeling side is in how black hole formation, growth, and feedback operate, and on modelling of dust attenuation and emission. 

However, these past conclusions have mainly been reached via comparison of the theoretical models and simulations with observations in the ``theoretical plane", i.e., by comparing the predicted physical properties such as stellar mass and SFR from simulations with \emph{estimates} of these physical properties obtained from SED fitting to observations. This approach has many advantages, including greater ease of interpretation in terms of intuitive physical quantities, and greater ease in linking populations across different epochs. However, it is very important to keep in mind that these estimates still carry significant uncertainties \citep[see e.g.][]{conroy:2013,leja:2017,leja:2019a}. Moreover, because of the complexity of the procedure used to obtain estimates of these derived quantities, it is very difficult to accurately and completely quantify their uncertainties, which are needed for a rigorous statistical assessment of the ``goodness of fit" of any theoretical model. The error budget should include contributions from the systematic uncertainties in deriving physical quantities such as stellar mass or SFR from SED fitting, as well as errors due to photometric redshift errors, photometric noise, and field-to-field variance. This detailed error budget has not been computed for quantities commonly used to calibrate models, such as stellar mass functions and SFR densities. 

We can get a first order sense for the possible systematic errors in the estimates of physical quantities by comparing the results from different methods. For stellar mass estimates using the {\sc zphot} code compared with the P12 method, which incorporates a more sophisticated prior on star formation history, we find systematic differences between the two methods of $0.2-0.5$ dex, and a dispersion of $0.1-0.5$ dex, with dependencies on stellar mass and redshift (similar to the findings of \citet{leja:2019a}). For SFR estimates, we find systematic errors for the P12 estimates compared with the \citet{barro:2019} estimates of typically at least 0.3-0.5 dex up to 2 dex, and a scatter of 0.5-1 dex. Both the systematic and random differences show dependencies on stellar mass and redshift. The {\sc zphot}-based SFR estimates show even larger differences compared with the \citet{barro:2019} SFR estimates, with different dependencies on stellar mass than the P12 SFR estimates. Clearly, SFR estimates from either rest-UV and available IR photometry and different SED fitting approaches still show very significant discrepancies which must be better understood. 

An alternative approach is to \emph{forward model} the simulations into the observational plane. This of course requires additional modeling steps, and the inclusion of assumptions regarding additional ingredients such as stellar population models and dust. However, estimating physical properties from SED fitting also contains similar assumptions, and in some cases one has more information about the conditions in the simulated galaxies than one does for the real galaxies. An additional advantage to working in the observational plane is that it is much easier to include modeling of observational errors and selection effects in this plane. We advocate carrying out comparison in both planes (theoretical and observational), as any differences in conclusions may illuminate problems. One of the important results of this work is that a comparison between theoretical predictions and observations in the ``theoretical plane" of stellar mass or SFR distribution functions versus the ``observational plane" of rest-UV and V-band luminosity functions appears to yield quantitatively different assessments of the goodness of fit of models compared with observations\footnote{We note that it is currently impossible to make rigorous statements about model goodness of fit due to the unavailability of complete, accurate error budgets, as discussed above. It may be that if the uncertainties on the observational quantities being used as constraints were properly accounted for in both cases, this difference would not be present.}. 

One of our main long term goals is to work towards a full forward modeling pipeline for multi-wavelength galaxy surveys. Over the next decade, wide area surveys from DESI, VRO, Euclid, the Nancy Grace Roman Space Telescope, 4MOST, and other facilities will be carried out. We can use the legacy observations from surveys such as CANDELS, to build a foundation for interpreting these new surveys. What we have shown here is that the current generation of semi-analytic models produce decent broad agreement with key properties of galaxy evolution as represented by CANDELS over the redshift range $0.5 \lesssim z \lesssim 3$. It has been shown elsewhere that these models produce similar results to those of numerical cosmological simulations and other semi-analytic models \citep{SD15}, and that they are also in agreement with higher redshift observations of galaxy populations \citep{Yung:2019a,Yung:2019b}, the reionization history \citep{Yung:2020}, and observational probes of the cold gas phase in galaxies \citep{Popping:2014,Popping:2019}. While there are certainly remaining tensions with observations, as seen here and also in e.g. \citet{Popping:2019}, there is promising ongoing work to continue to improve the realism of the treatment of physical processes in SAMs \citep[e.g.][]{Pandya:2020}. In work in progress, we are using this framework to create similar mock observations for future planned surveys with the James Webb Space Telescope and the Nancy Grace Roman Space Telescope (L. Y. A. Yung et al. in prep).  SAMs coupled with lightcones extracted from large volume N-body simulations have recently been used to create a 2 sq. deg. lightcone from $0 < z< 10$ \citep[][Yung et al. in prep]{yang:2020}. In order to create mock surveys for even larger areas --- tens to hundreds of square degrees --- that will be probed by the projects mentioned above, it is likely that new, even more computationally efficient techniques will need to be developed, perhaps enabled by machine learning based tools.

\section{Conclusions}
\label{sec:conclusions}

In this paper, we presented mock lightcones that were custom created to aid in the interpretation of observations from the CANDELS program. We populated these lightcones with galaxies using two different semi-analytic modeling codes, and the empirical model {\sc UniverseMachine}. In addition, we presented specially curated ``theory friendly" catalogs for the CANDELS observations, which include a selection of the observed and rest-frame photometry as well as estimates of physical galaxy properties such as redshift, stellar mass, and star formation rate. We make all data products available through a web-based data hub that allows users to preview and download the data. 

We showed comparisons between the mock lightcones and the CANDELS observations for a selection of key quantities in the ``observational plane", including observed frame counts, rest-frame luminosity functions, color-magnitude and color-color distributions. We also compared our model predictions with physical quantities estimated via SED-fitting from the CANDELS photometry, such as stellar mass functions, SFR distribution functions, the stellar mass vs. SFR relation, and dust attenuation. Although there are some tensions between the theoretical predictions and the observations, we conclude that these mock catalogs reproduce the observational estimates accurately enough to be useful for interpreting current observations and making predictions for future ones. 

\section*{Acknowledgements}
We thank the anonymous referee for helpful comments that improved the manuscript. We thank the Flatiron Institute for providing computing resources and data access. We warmly thank Dylan Simon, Elizabeth Lovero, and Austen Gabrielpillai for building Flathub. We thank Yotam Cohen for useful comments on the Flathub datahub. RSS is supported by the Simons Foundation. CP is supported by the Canadian Space Agency under a contract with NRC Herzberg Astronomy and Astrophysics. This work makes use of observations taken by the CANDELS Multi-Cycle Treasury Program with the NASA/ESA HST, which is operated by the Association
of Universities for Research in Astronomy, Inc., under NASA contract NAS5-26555. This work is based in part on observations made with the Spitzer Space Telescope, which was operated by the Jet Propulsion Laboratory, California Institute of Technology under a contract with NASA. 

\section*{Data Availability}
The data underlying this article are available from the Flatiron Institute Data Exploration and Comparison Hub (Flathub), at \url{http://flathub.flatironinstitute.org/group/candels}.

\appendix
\section{Supplementary Figures}
\label{sec:appendix}
\begin{figure*} 
\begin{center}
\includegraphics[width=0.8\textwidth]{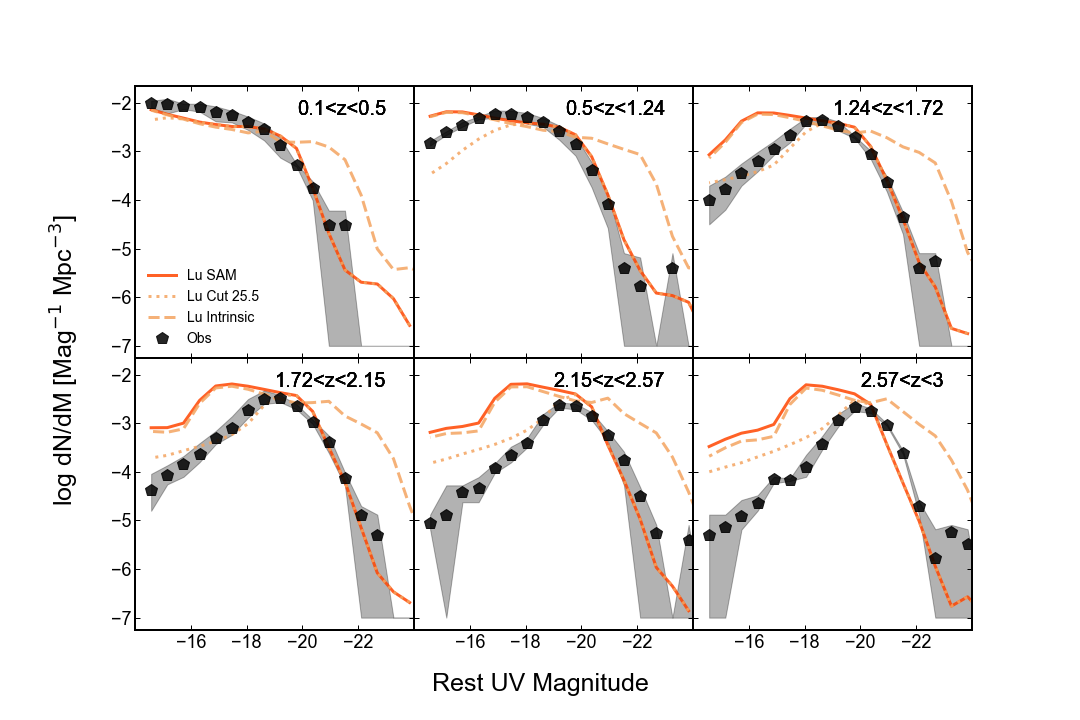}
\end{center}
\caption{Luminosity functions in the rest-frame UV ($1500$ \AA) divided into redshift bins, for
  the Lu-SAM, compared with the corresponding distribution from CANDELS (black symbols show the mean over all fields; shaded areas show the minimum and maximum value in each bin over the four fields). Dashed lines show the intrinsic luminosity functions with no dust attenuation; solid lines show the model predictions with dust attenuation included. Dotted lines show the dust attenuated model predictions with a cut of F160W$<25.5$ applied (similar to the observations). The Lu SAM predictions agree with the observed distributions quite well in the regime where the observations are highly complete. 
\label{fig:lf_UV_Lu}}
\end{figure*}

\begin{figure*} 
\begin{center}
\includegraphics[width=0.8\textwidth]{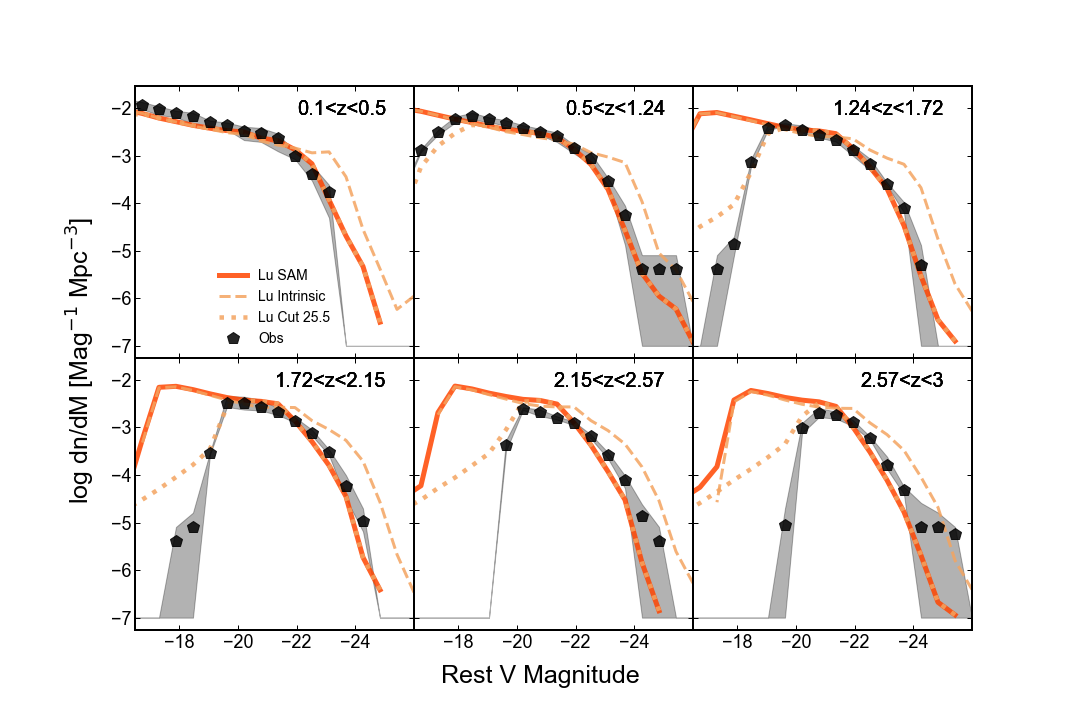}
\end{center}
\caption{Luminosity functions in the rest-frame V-band divided into
  redshift bins for the Lu-SAM and CANDELS observations. Key is as in Fig.~\ref{fig:lf_UV_Lu}. The Lu SAM predictions agree with the observed rest-frame V-band magnitude distributions fairly well in the regime where the observations are highly complete.
\label{fig:lf_V_Lu}}
\end{figure*}

\begin{figure*} 
\begin{center}
\includegraphics[width=0.8\textwidth]{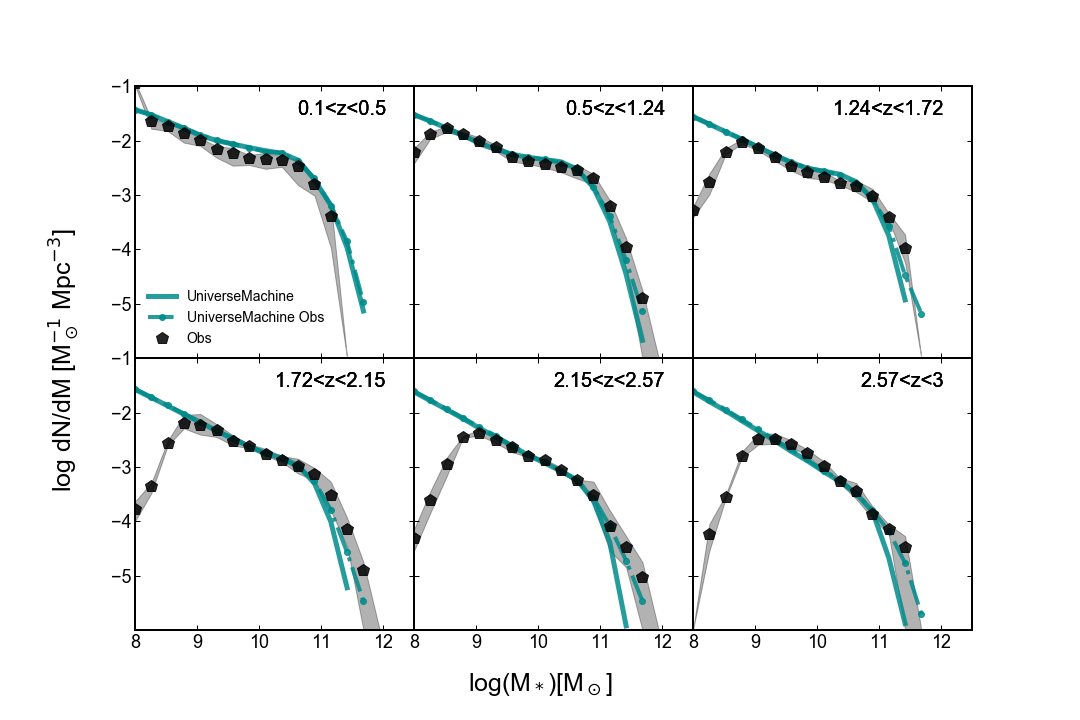}
\end{center}
\caption{Stellar mass functions divided into redshift bins, for {\sc UniverseMachine}, compared with stellar mass distribution functions derived from CANDELS. Solid lines show the results for the intrinsic (error-free) stellar mass predictions in {\sc UniverseMachine}, while dot-dashed lines show the predictions after modeling the expected errors on the stellar masses. Errors in stellar mass estimates can be significant, especially at high redshift, and lead to an Eddington bias that impacts the high-mass end of the distribution. 
\label{fig:SMF_obs}}
\end{figure*}

\begin{figure*} 
\begin{center}
\includegraphics[width=0.8\textwidth]{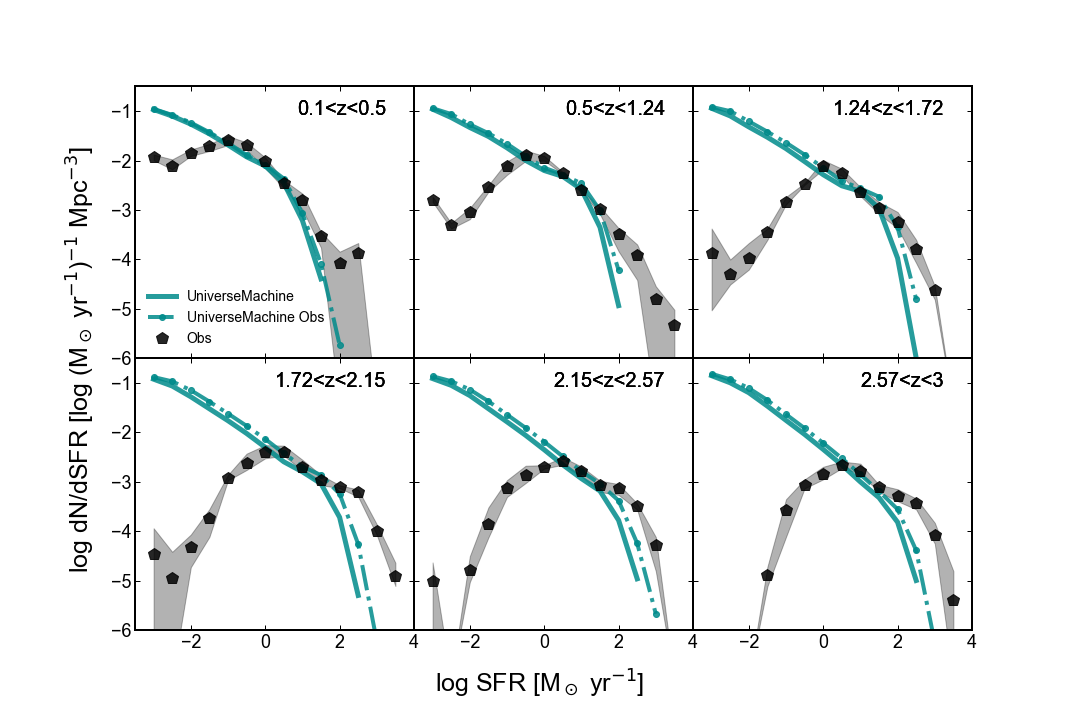}
\end{center}
\caption{Star formation rate functions divided into redshift bins, for {\sc UniverseMachine}, compared with SFR functions derived from CANDELS. Solid lines show the results for the intrinsic (error-free) SFR predictions in {\sc UniverseMachine}, while dot-dashed lines show the predictions after modeling the expected errors on the SFR estimates.
\label{fig:SFRF_obs}}
\end{figure*}

In this appendix we show results that supplement those in the main text. Fig.~\ref{fig:lf_UV_Lu} and \ref{fig:lf_V_Lu} show the rest-UV and V-band luminosity functions for the Lu SAMs compared with the CANDELS observations. Fig.~\ref{fig:SMF_obs} and \ref{fig:SFRF_obs} show the stellar mass function and SFRF from {\sc UniverseMachine} with observational errors included. For details please see the main text. 

\bibliographystyle{mn}
\bibliography{mn-jour,candelslc}

\end{document}